\documentclass[]{piparticle-final}
\usepackage{graphicx}
\usepackage{amssymb}
\usepackage{amsmath}
\usepackage{cite}

\begin{document}
\volume{3}               
\articlenumber{030004}   
\journalyear{2011}       
\editor{J-C. G\'eminard}   
\reviewers{B. Tighe, Instituut-Lorentz, Universiteit Leiden, Netherlands.}  
\received{$11$ May 2011}     
\accepted{15 July 2011}   
\runningauthor{L. A. Pugnaloni \itshape{et al.}}  
\doi{030004}         

\title{Master curves for the stress tensor invariants in stationary states of static granular beds. Implications for the thermodynamic phase space}

\author{Luis A. Pugnaloni,\cite{inst1}\thanks{E-mail: luis@iflysib.unlp.edu.ar}\hspace{.5em}
        Jos\'e Damas,\cite{inst2}\thanks{E-mail: jdelacruz@alumni.unav.es} \hspace{.5em}
        Iker Zuriguel,\cite{inst2}\thanks{E-mail: iker@unav.es} \hspace{.5em}
        Diego Maza\cite{inst2}\thanks{E-mail: dmaza@unav.es}
	}

\pipabstract{
We prepare static granular beds under gravity in different stationary states by tapping the system with pulsed excitations of controlled amplitude and duration. The macroscopic state---defined by the ensemble of static configurations explored by the system tap after tap---for a given tap intensity and duration is studied in terms of volume, $V$, and force moment tensor, $\Sigma$. In a previous paper [Pugnaloni et al., Phys. Rev. E  82, 050301(R) (2010)], we reported evidence supporting that such macroscopic states cannot be fully described by using only $V$ or $\Sigma$, apart from the number of particles $N$. In this work, we present an analysis of the fluctuations of these variables that indicates that $V$ and $\Sigma$ may be sufficient to define the macroscopic states. Moreover, we show that only one of the invariants of $\Sigma$ is necessary, since each component of $\Sigma$ falls onto a master curve when plotted as a function of $\rm{Tr}(\Sigma)$. This  implies that these granular assemblies have a common shape for the stress tensor, even though it does not correspond to the hydrostatic type. Although most results are obtained by molecular dynamics simulations, we present supporting experimental results.}
\maketitle

\blfootnote{
\begin{theaffiliation}{99}
   \institution{inst1} Instituto de F\'{\i}sica de L\'{\i}quidos y Sistemas Biol\'ogicos, (CONICET La Plata, UNLP), Calle 59 No 789, 1900 La Plata, Argentina.
    \institution{inst2} Departamento de F\'{\i}sica y Matem\'atica Aplicada, Facultad de Ciencias, Universidad de Navarra, Pamplona, Spain.
\end{theaffiliation}
}

\section{Introduction}
The study of static granular systems is of fundamental importance to the industry to improve the storage of such materials in bulk, as well as to optimize the packaging design. However, far from yielding such benefits of practical interest, physicists have found a fascinating challenge on their way that has stuck them, to a large extent, in the area of static granular matter. Such challenge is finding an appropriate description of the simplest state in which matter can be found: \textit{equilibrium} \cite{callen}. At equilibrium, a sample will explore different microscopic configurations over time, in such a way that macroscopic averages over large periods will be well defined, with no aging. Moreover, if not only equilibrium but ergodicity is present, averages over a large number of replicas at a given time should give equivalent results to time averages \cite{pathria}. Finally, one expects that all macroscopic properties of such equilibrium states can be put in terms of a few independent macroscopic variables. Then, a thermodynamic description and, hopefully, a statistical mechanics approach can be attempted. Theoretical formalisms based on these assumptions have been used to analyze data from experiments and numerical models. However, some of the foundations are still supported by little evidence.

The use of careful protocols to make a granular sample explore microscopic configurations within a seemingly equilibrium macroscopic state has given us the first standpoint \cite{nowak,richard,schroter,ribiere}. In such protocols, the sample is subjected to external excitations of the form of pulses. Well defined, reproducible time averages are found after a transient if the same pulse shape and pulse intensity are applied. However, for low intensity pulses, a previous annealing might be in order since equilibrium is hard to reach ---as in supercooled liquids. We will use the expressions \textit{equilibrium state}, \textit{steady state} or simply \textit{state} to refer to the collection of all configurations generated by using a given external excitation after any transient has disappeared.

More than twenty years ago, Edwards and Oakshott \cite{edwards} put forward the idea that the number of grains $N$ and the volume $V$ are the basic state variables that suffice to characterize a static sample of hard grains in equilibrium. The $NV$ granular ensemble was then introduced as a collection of microstates, where the sample is in mechanical equilibrium, compatible with $N$ and $V$. However, newer theoretical works \cite{snoeijer,edwards2,blumenfeld,henkes,henkes2,tighe} suggest that the force moment tensor, $\Sigma$, ($\Sigma=V \sigma$, where $\sigma$ is the stress tensor) must be added to the set of extensive macroscopic variables (i.e., an $NV\Sigma$ ensemble) to adequately describe a packing of real grains.

In the rest of this paper, we show experimental and simulation evidence that the equilibrium states of static granular packings cannot be only described by $V$ (or equivalently the packing fraction, $\phi$, defined as the fraction of the space covered by the grains) nor by $\Sigma$. We do this by generating states of equal $V$ but different $\Sigma$ and states of equal $\Sigma$ but different $\phi$. We also show that states of equal $V$ may present different volume fluctuations. Moreover, we show that states of equal $V$ and $\Sigma$ display the same fluctuations of these variables, suggesting that no other extensive parameter might be required to characterize the state (apart from $N$). Finally, but of major significance, we show that the shape of the force moment tensor is universal, in the sense that different states that present the same trace of the tensor actually have the same value in all the components of $\Sigma$.

\section{Simulation}

We use soft-particle 2D molecular dynamics \cite{pugnaloni2008,arevalo}. Particle--particle interactions are controlled by the particle--particle overlap $\xi =d-\left\vert \mathbf{r}_{ij}\right\vert $ and the velocities $\dot{\mathbf{r}}_{ij}$, $\omega _{i}$ and $\omega _{j}$. Here, $\mathbf{r}
_{ij}$ represents the center-to-center vector between particles $i$ and $j$, $d$ is the particle diameter and $\omega $ is the particle angular velocity. These forces are introduced in the Newton's translational and rotational equations of motion and then numerically integrated by a velocity Verlet algorithm \cite{Schafer1}. The interaction of the particles with the flat surfaces of the container is calculated as the interaction with a disk of infinite radius. 

The contact interactions involve a normal force $F_{\text{n}}$ and a tangential force $F_{\text{t}}$. 

\begin{equation}
F_{\text{n}}=k_{\text{n}}\xi -\gamma _{\text{n}}v_{i,j}^{\text{n}}
\label{normal}
\end{equation}

\begin{equation}
F_{\text{t}}=-\min \left( \mu |F_{\text{n}}|,|F_{\text{s}}|\right) \cdot 
\text{sign}\left( \zeta \right)  \label{tangent}
\end{equation}%
where

\begin{equation}
F_{\text{s}}=-k_{\text{s}}\zeta -\gamma _{\text{s}}v_{i,j}^{\text{t}}
\label{contact}
\end{equation}

\begin{equation}
\zeta \left( t\right) =\int_{t_{0}}^{t}v_{i,j}^{\text{t}}\left( t^{\prime
}\right) dt^{\prime }  \label{static}
\end{equation}

\begin{equation}
v_{i,j}^{\text{t}}=\dot{\mathbf{r}}_{ij}\cdot \mathbf{s}+\frac{1}{2}d\left(
\omega _{i}+\omega _{j}\right)  \label{vtan}
\end{equation}

The first term in Eq.~(\ref{normal}) corresponds to a restoring force proportional to the superposition $\xi $ of the interacting disks and the stiffness constant $k_{n}$. The second term accounts for the dissipation of energy during the contact and is proportional to the normal component $v_{i,j}^{\text{n}}$ of the relative velocity $\dot{\mathbf{r}}_{ij}$ of the
disks.

Equation~(\ref{tangent}) provides the magnitude of the force in the tangential direction. It implements the Coulomb's criterion with an effective friction following a rule that selects between static or dynamic friction. Notice that Eq. (\ref{tangent}) implies that the maximum static friction force $|F_{\text{s}}|$ used corresponds to $\mu |F_{\text{n}}|$, which effectively sets $\mu_{\text{dynamic}}=\mu_{\text{static}}=\mu$. The static friction force $F_{\text{s}}$ [see Eq. (\ref{contact})] has an
elastic term proportional to the relative shear displacement $\zeta $ and a dissipative term proportional to the tangential component $v_{i,j}^{\text{t}} $ of the relative velocity. In Eq. (\ref{vtan}), $\mathbf{s}$ is a unit vector normal to $\mathbf{r}_{ij}$. The elastic and dissipative contributions are characterized by $k_{\text{s}}$ and $\gamma _{\text{s}}$, respectively. The shear displacement $\zeta $ is calculated through Eq. (\ref{static}) by integrating $v_{i,j}^{\text{t}}$ from the beginning of the contact (i.e., $t=t_{0}$). The tangential interaction behaves like a damped spring which is formed whenever two grains come into contact and is removed when the contact finishes \cite{wolf}. 

The particular set of parameters used for the simulation is (unless otherwise stated): $\mu = 0.5$, $k_n = 10^5 (mg/d)$, $\gamma_n = 300 (m\sqrt{g/d})$, $k_s= \frac{2}{7}k_n$ and $\gamma_s = 200 (m\sqrt{g/d})$. In some cases, we have varied $\mu$ and $\gamma_n$ in order to control the friction and restitution coefficient. The integration time step is set to $\delta = 10^{-4} \sqrt{d/g}$. The confining box ($13.39d$-wide and infinitely high) contains $N=512$ monosized disks. Units are reduced with the diameter of the disks, $d$, the disk mass, $m$, and the acceleration of gravity, $g$. 

Tapping is simulated by moving the confining box in the vertical direction following a half sine wave trajectory [$A\sin(2\pi\nu t)(1-\Theta(2\pi\nu t-\pi))$]. The excitation can be controlled through the amplitude, $A$, and the frequency, $\nu$, of the sinusoidal trajectory. We implement a robust criterion based on the stability of particle contacts to decide when the system has reached mechanical equilibrium \cite{arevalo} before a new tap is applied to the sample. Averages were taken over 100 taps in the steady state and over 20 independent simulations for each value of $A$ and $\nu$.

The volume, $V$, of the system after each tap can be obtained from the packing fraction, $\phi$, as $V=N\pi(d/2)^2/\phi$. We measure $\phi$ in a rectangular window centered in the center of mass of the packing. The measuring region covers 90\% of the height of the granular bed (which is of about $40d$) and avoids the area close to the walls by $1.5d$. We have observed that $\phi$ is sensitive to the chosen window. However, none of the conclusions drawn in this paper are affected by this choice. 

The stress tensor, $\sigma$, is calculated from the particle--particle contact forces as

\begin{equation}
\sigma^{\alpha\beta} = \frac{1}{V} \sum_{c_{ij}}{r^{\alpha}_{ij}f^{\beta}_{ij}},
\end{equation}
where the sum runs over all contacts.

The force moment tensor, $\Sigma$, is defined as $\Sigma \equiv V\sigma$. During the course of a tap $\Sigma$ is non-symmetric, however, once mechanical equilibrium is reached in accordance with our criterion, $\Sigma$ becomes symmetric within a very small error compared with the fluctuations of $\Sigma$. Although $\Sigma$ may depend on depth, we have measured the force moment tensor by simply summing over particle--particle contacts in the entire system.

The fluctuations of $\phi$ ($\Delta \phi$) and $\Sigma$ ($\Delta \Sigma$) are calculated as the standard deviation in the 100 taps obtained in each steady state. We average $\phi$, $\Sigma$, and their fluctuations over 20 independent runs for each steady state and estimate error bars as the standard deviation over these 20 runs.

\section{Experimental method}

\begin{figure}[htp]
\begin{center}
 \includegraphics[width=0.48\textwidth,angle=0]{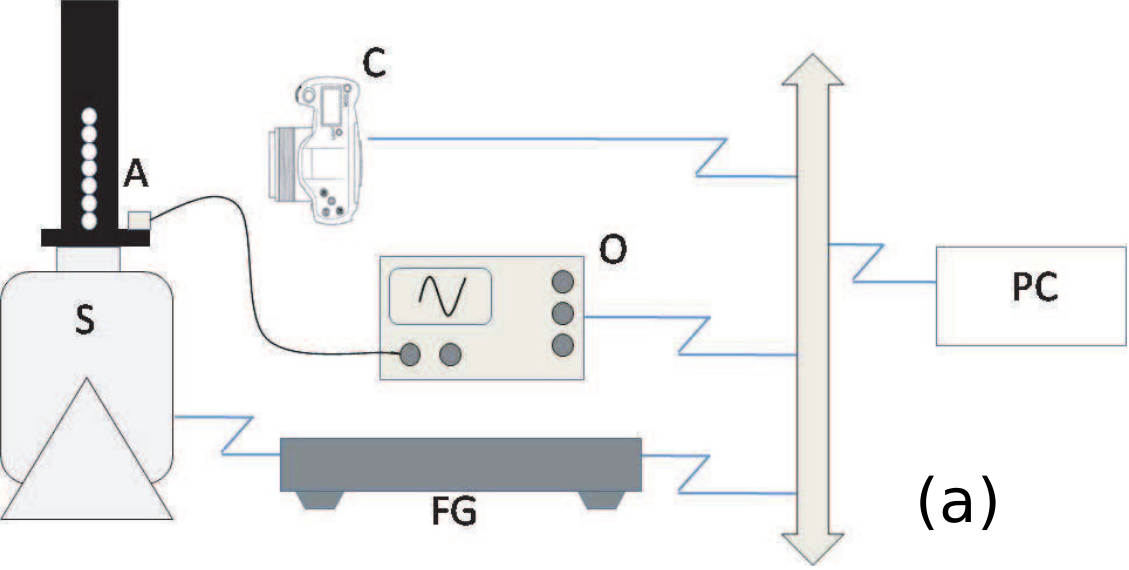}
 \includegraphics[width=0.28\textwidth,angle=0]{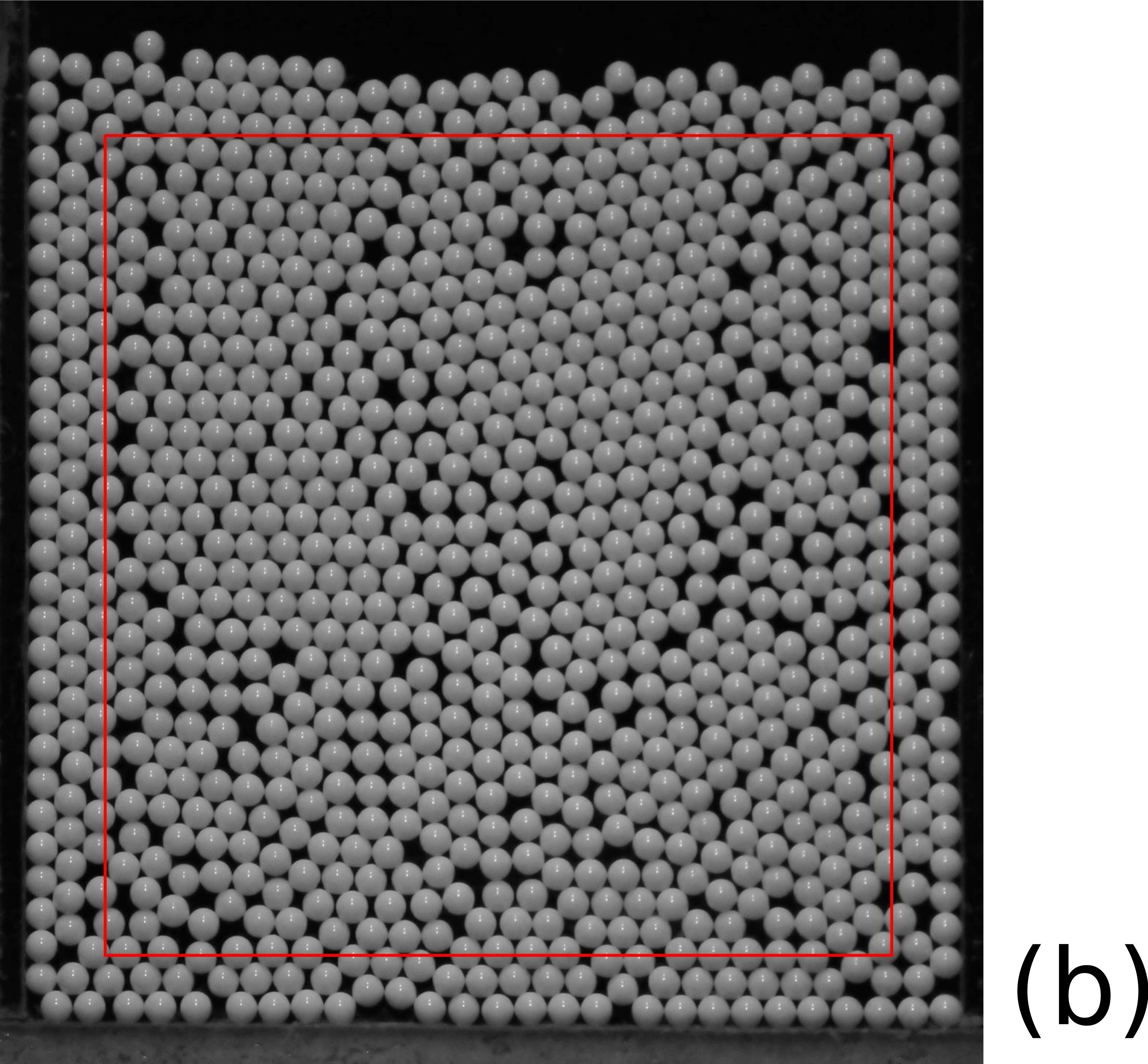}
\end{center}
 \caption{(a) Schematic diagram of the experimental set-up. A: accelerometer, S: shaker, C: camera, O: oscilloscope, FG: function generator, PC: computer. (b) Example of an image of the packing. The region of measurement is indicated by a red square.}
 \label{fig:sketch_expt}
\end{figure}

The experimental set-up is sketched in Fig. \ref{fig:sketch_expt}(a). A quasi 2D Plexiglass cell (28 mm wide and 150 mm high) is filled with 900 alumina oxide beads of diameter $d=1\pm0.005$ mm. The separation between the front and rear plates was made 15\% larger than the bead diameter. The cell is tapped by an electromagnetic shaker (Tiravib 52100) with a trend of half sine wave pulses separated five seconds between them. The tapping amplitude was controlled adjusting the intensity, $A$, and frequency, $\nu$, of the pulse and was measured by an accelerometer attached to the base of the cell. Averages are taken over 500 taps after equilibration.

High resolution images (more than 10 MPix) are taken after each tap. To calculate the packing fraction, we only consider a rectangular zone at the center of the packing whose limits are at 4 mm from the borders [see Fig. \ref{fig:sketch_expt}(b)]. We determine the centroid of each particle by means of a numerical algorithm with subpixel resolution. Then, we calculate the packing fraction by assuming that the 2D projection of each bead corresponds to a disk of diameter $d=1$ mm. We estimate the packing fraction with a resolution of $\pm 0.001$. As the separation between plates is larger than the particle diameter, small overlaps between the spheres are present in the 2D projections of the images. Therefore, the calculated packing fraction in some dense configurations might result slightly higher than the hexagonal disk packing limit ($\pi\sqrt{3}/6$).

\section{Tapping characterization and asymptotic equilibrium states}

It is often debated \cite{dijksman,ludewing} what is the appropriate parameter to characterize the external excitation used to drive a granular sample. Dijksman et al. \cite{dijksman} proposed a parameter related to the lift-off velocity of the granular bed. Ludewing et al. \cite{ludewing} presented an energy based parameter. Pugnaloni et al. \cite{pugnaloni2008, gago} suggested that the factor of expansion induced on the sample would be a suitable measure \cite{expansion}. The usual perspective to define a pulse parameter, $\epsilon$, is to achieve a collapse of the $\phi$--$\epsilon$ curves as different details of the pulse shape are changed (such as amplitude and duration). Parameters defined in all these previous works fail to make such curves collapse for the data presented in this paper. The main reason for this is that our $\phi$--$\epsilon$ curves are non-monotonic (see for example Fig. \ref{fig:phi-vs-Gamma_sim}) presenting a minimum whose depth depends on the details of the pulse shape. Therefore, a simple rescale of the horizontal axis does not suffice to collapse the curves. 

Since we are interested in macroscopic states, the actual pulse used to drive the system is merely a control parameter but not a macroscopic variable that describes the state. Therefore, the external pulse does not need to be described with a simplified quantity. The complete functional form of the pulse can be given instead. In our case, we use a sine pulse and both, the pulse amplitude and frequency, are needed to fully describe the excitation. We will employ the usual nondimensional peak acceleration $\Gamma \equiv a_{peak}/g = A (2\pi\nu)^2/g$ (where $g$ is the acceleration of gravity) and the frequency $\nu$ to precisely define the external excitation. A detailed study of the dynamical response of a granular bed to a pulse of controlled intensity and duration can be found in Damas et al. \cite{damas}.

One major issue in studying equilibrium states is the evidence that one can have, indicating if the system is actually at equilibrium \cite{ribiere}. Since the definition of equilibrium is circular \cite{equilibrium}, we can simply do our best to check if different properties of the system have well defined means (as well as higher order moments of the distributions) which should not depend on the history of the processes applied to the sample.

\begin{figure}[htp]
\begin{center}
 \includegraphics[width=0.4\textwidth,angle=0]{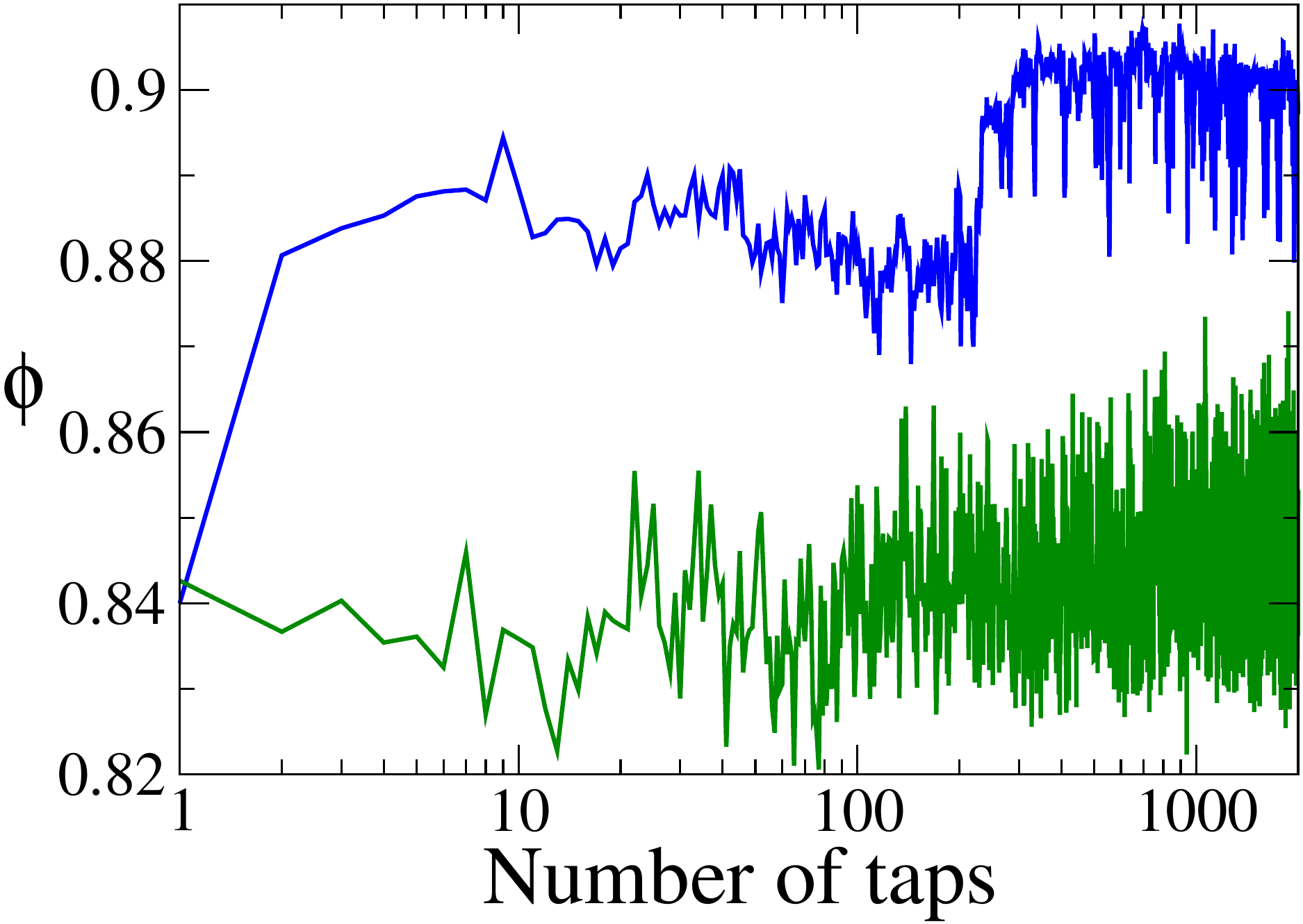}
\end{center}
 \caption{Evolution of $\phi$ towards the steady state starting from a disordered configuration initially obtained by strong taps (experimental results). Results for two tapping intensities are shown: $\Gamma = 4.8$ (blue), and $\Gamma = 17$ (green). In both experiments, the frequency of the pulse is $\nu=30$ Hz.}
 \label{fig:history_expt}
\end{figure}

We have generated configurations corresponding to a particular pulse (of a given shape and intensity) by repeatedly applying such pulse to the system and allowing enough time for any transient to fade. To prove that our samples are at equilibrium, we approach a particular pulse through different paths ---and starting from different initial conditions (ordered and disordered configurations)--- and confirm that the steady states obtained present equivalent mean values and second moments of the distributions of the variables of interest.

In Ref. \cite{pugnaloni2010}, we showed that the steady states corresponding to excitations of high intensity are reached in a few taps, even if the initial configuration corresponds to a very ordered structure. In Fig. \ref{fig:history_expt}, we consider the equilibration process from a highly disordered initial configuration. For low tapping intensities (blue line in Fig. \ref{fig:history_expt}), the packing fraction evolves to the steady state in two stages. Just after switching the tapping amplitude to the reference value, the system rapidly evolves to values of  $\phi$ close to the final steady state. Beyond this initial convergence, a slower compaction phase takes the system to the final steady state. For high tapping intensities, the evolution to the steady state is very rapid. The steady state is reached after about a hundred taps [green line in Fig. \ref{fig:history_expt})]. Therefore, we apply a sequence of at least 1000 taps in all our experiments before taking averages to warrant that the steady state has been reached. In our simulations, 400 taps of equilibration were enough.

\begin{figure}[htp]
\begin{center}
 \includegraphics[width=0.4\textwidth,angle=0]{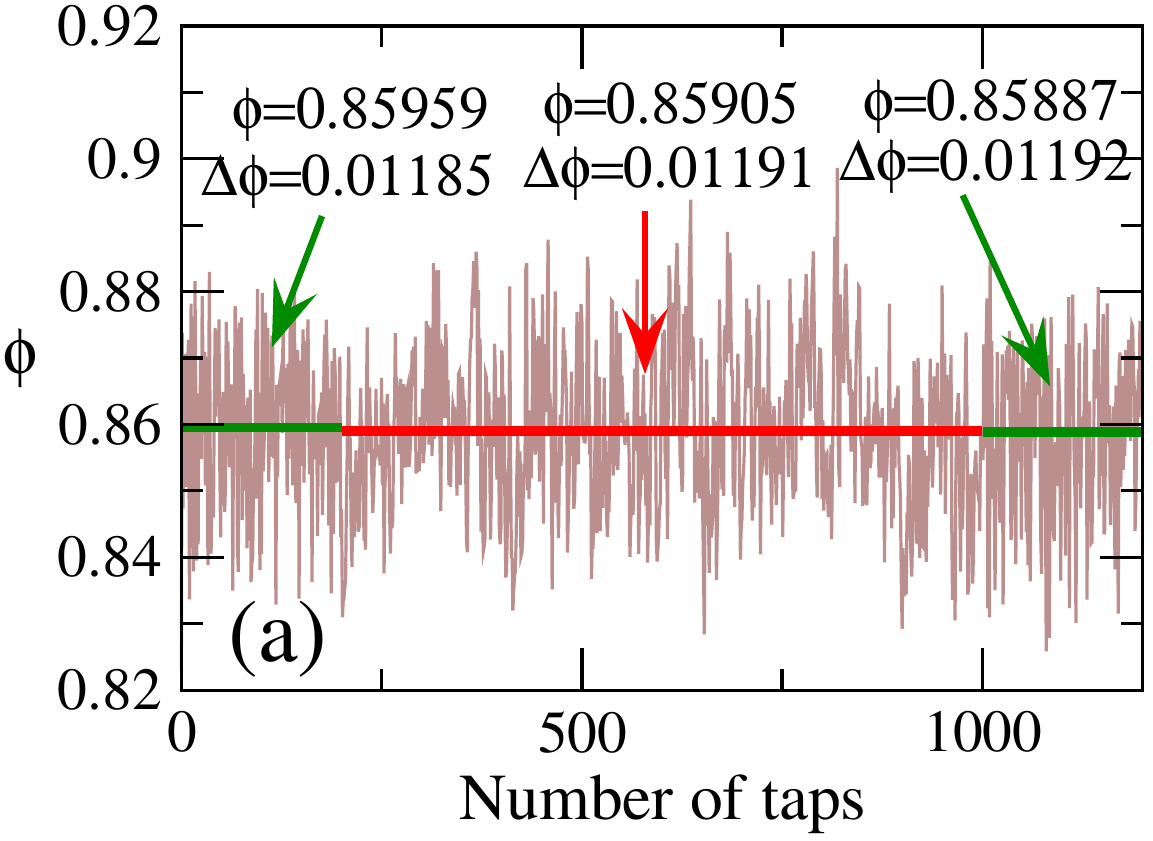}
 \includegraphics[width=0.4\textwidth,angle=0]{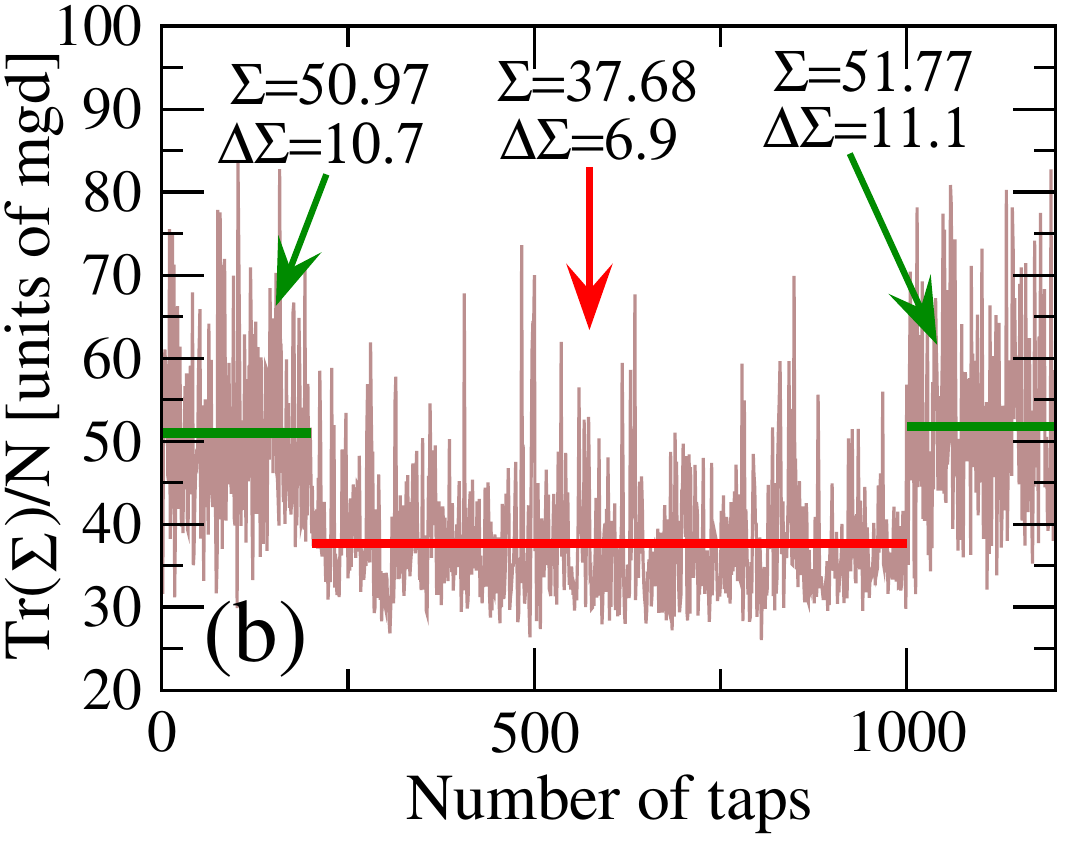}
\end{center}
 \caption{Evolution of $\phi$ (a) and $\rm{Tr}(\Sigma)$ (b) as the pulse intensity is suddenly changed between two values that produce steady states with the same $\phi$ but different $\Sigma$ in our simulations. The initial 200 taps correspond to $\Gamma=4.9$, the middle 800 pulses to $\Gamma=61.5$ and the final 200 pulses, again, to $\Gamma=4.9$. In all cases, $\nu=0.5 \sqrt{g/d}$. The mean values and standard deviations in each section are indicated by arrows.}
 \label{fig:history_simul}
\end{figure}

An interesting example of equilibration is presented in Fig. \ref{fig:history_simul}. In this case, we present the values of $\phi$ and of $\rm{Tr}(\Sigma)$ during a very special sequence of pulses obtained in our simulations. The system is initially deposited from a dilute random configuration with all particles in the air. First, we tap the system 200 times at $\Gamma=4.9$, then 800 times at $\Gamma=61.5$ and finally, 200 times at $\Gamma=4.9$. In the whole run, we keep $\nu=0.5\sqrt{g/d}$. These two values of the pulse intensity have been chosen because they are known to produce packings with the same mean $\phi$ in the steady state \cite{pugnaloni2010}. However, the mean $\Sigma$ is clearly different. Notice, however, that the same values of $\phi$, $\Sigma$ and its fluctuations, $\Delta \phi$ and $\Delta \Sigma$, are observed for the steady state of low $\Gamma$ obtained before and after the 800 pulses of high $\Gamma$.

Unless otherwise stated, all the results we present in what follows correspond to steady states. We have tested this by obtaining the same states through two different preparation protocols consisting of: (i) application of a large number of identical pulses starting from a disordered configuration, and (ii) application of a reduced number of identical pulses after annealing the system from much higher pulse intensities. In a few cases, the results (mean values and/or fluctuations) from both protocols did not match. This was an indication that a steady state, if existed, was not reached by one of the protocols (or by both protocols). This happens especially for low intensity taps, which require longer equilibration times. Such cases have been removed from the analysis.

\section{Volume and volume fluctuations}

\begin{figure}[htp]
\begin{center}
 \includegraphics[width=0.4\textwidth,angle=0]{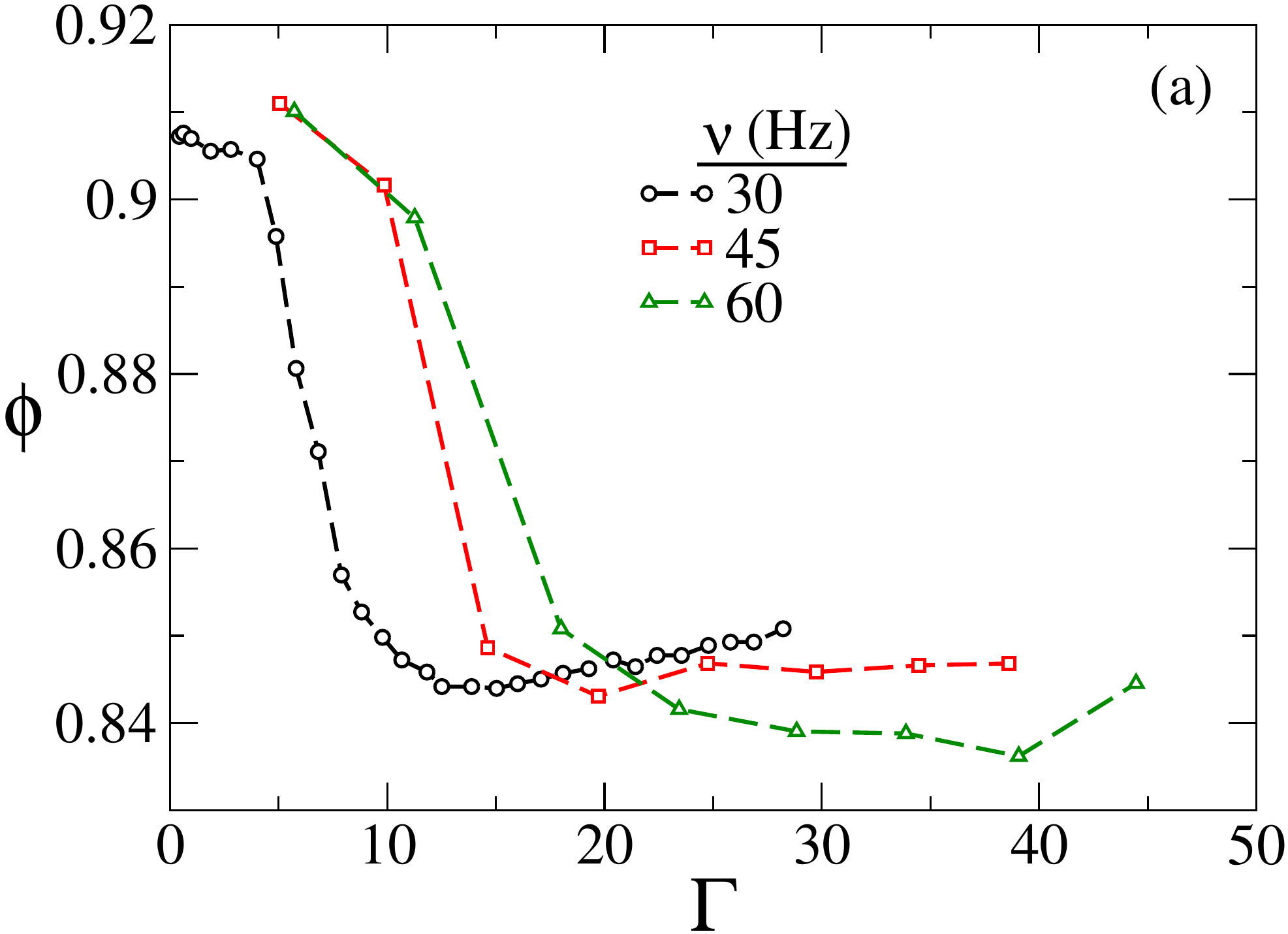}
 \includegraphics[width=0.41\textwidth,angle=0]{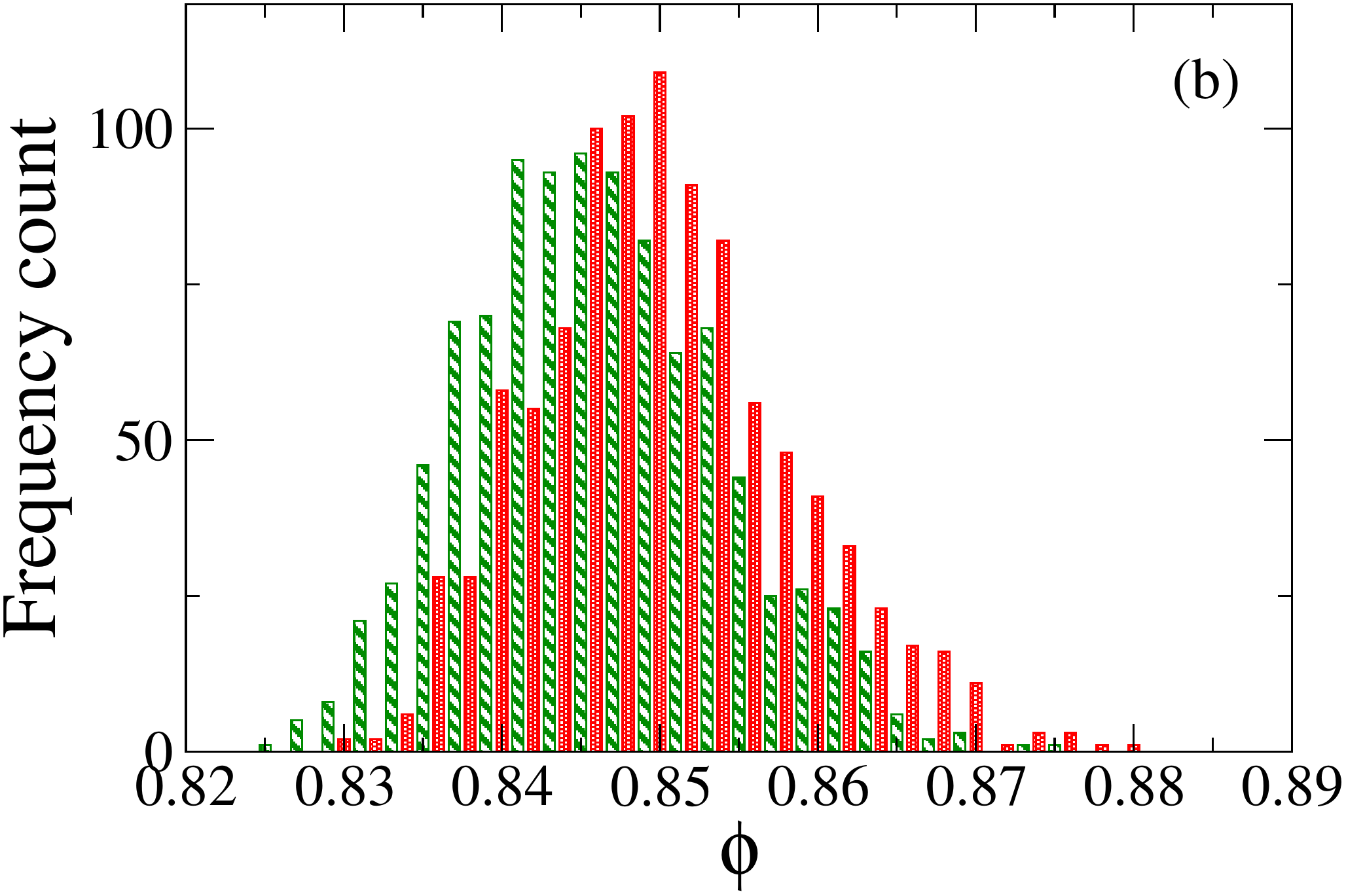}
 \end{center}
 \caption{(a) Experimental results for the steady state packing fraction, $\phi$, as a function of the reduced peak acceleration, $\Gamma$, for different frequencies, $\nu$, of the tap pulse. (b) Histogram of the configurations visited by the system for $\nu=30$ Hz: $\Gamma=15$ (green) and $\Gamma=28$ (red).}
 \label{fig:phi-vs-Gamma_expt}
\end{figure}

In Fig.~\ref{fig:phi-vs-Gamma_expt}(a), we plot $\phi$ in the steady state as a function of $\Gamma$ from our experiments for different pulse frequencies \cite{note1}. As we can see, there exists a minimum $\phi$ at relatively high $\Gamma$. A similar experiment on a three-dimensional cell also yielded analogous results \cite{pugnaloni2010}. This behavior has also been reported for various models \cite{pugnaloni2008,gago,carlevaro}. An explanation for this, based on the formation of arches, has been given in \cite{pugnaloni2008}. The position of the minimum in $\phi$ shifts to larger $\Gamma$ if the frequency of the pulse is increased (i.e., if the pulse duration is reduced). 

\begin{figure}[htp]
\begin{center}
 \includegraphics[width=0.41\textwidth,angle=0]{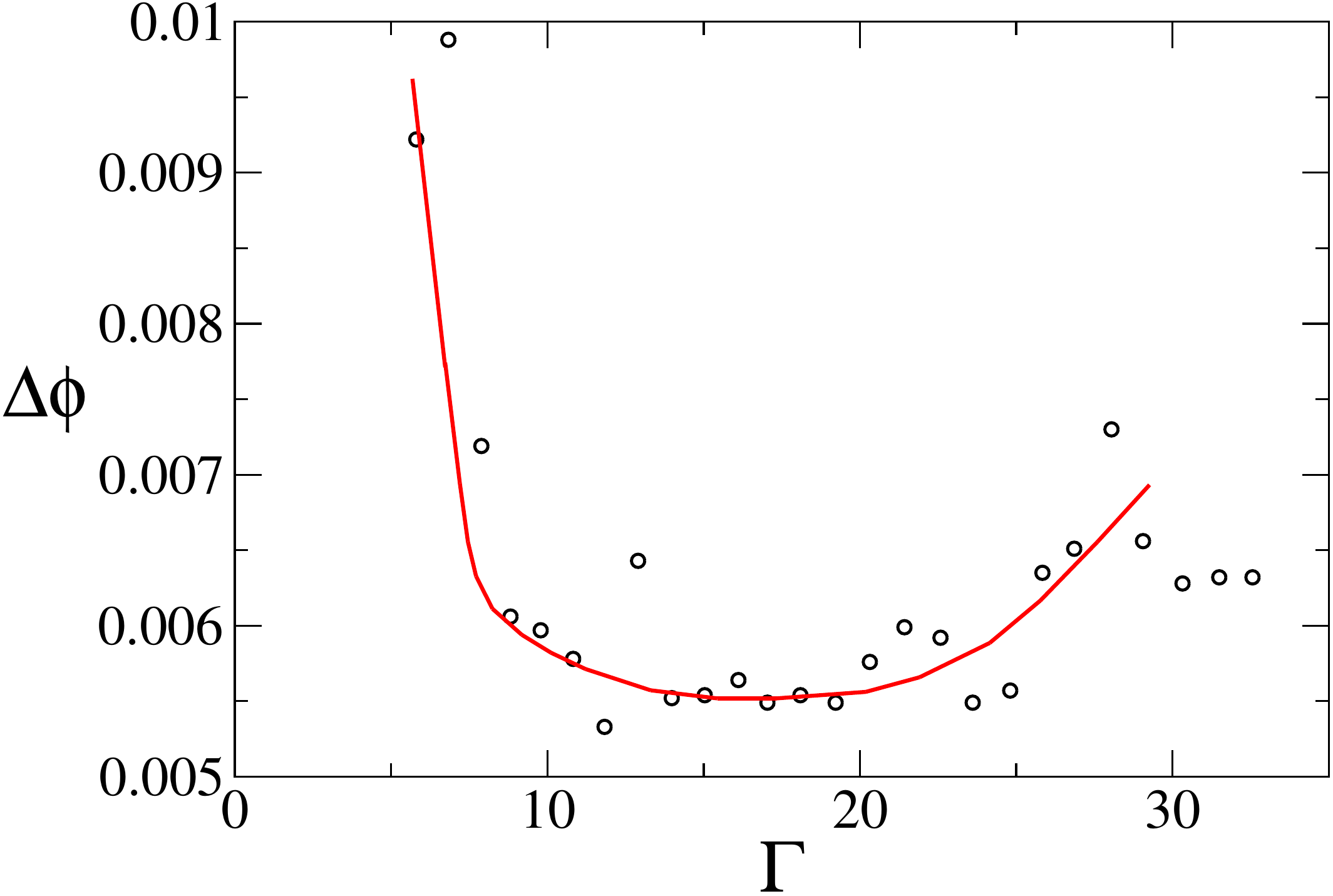}
\end{center}
 \caption{The steady state volume fraction fluctuations, $\Delta \phi$, as a function of $\Gamma$ from our experiments with $\nu=30$ Hz. The solid line is only a guide to the eyes.}
 \label{fig:D_phi-vs-Gamma_expt}
\end{figure}

Although the increment in the packing fraction beyond the minimum is rather small, it is important to remark that this difference is not an artifact introduced by our experimental resolution. In Fig. \ref{fig:phi-vs-Gamma_expt}(b), we show the histogram for the sequence of packings obtained for $\Gamma\simeq 15$  (the minimum packing fraction for $\nu=30$ Hz) and for $\Gamma=28$ (the largest excitation explored for $\nu=30$ Hz). Both steady states are statistically comparable; however, it is possible to distinguish different mean values.  

Since steady states of equal $\phi$ obtained at both sides of the $\phi$ minimum are generated via very different tap intensities, it is worth assessing if such states are, in fact, equivalent. This can be done by comparing the volume fluctuations of such states. A similar analysis was done in Ref.~\cite{ciamarra} for states generated with different pulse amplitude and duration in liquid fluidized beds.

The fluctuations of $\phi$ in the steady state, as measured by the standard deviation $\Delta \phi$, are presented in Fig.~\ref{fig:D_phi-vs-Gamma_expt}. As we can see, a minimum in the fluctuations is just apparent. The position of such minimum in the fluctuations coincides with the minimum in $\phi$. Unfortunately, the resolution of $\phi$ in our experiments is around the size of the fluctuations. However, the results from the simulations are not limited in this respect and we address to those for a more reliable assessment of the fluctuations.

\begin{figure}[htp]
\begin{center}
 \includegraphics[width=0.4\textwidth,angle=-0]{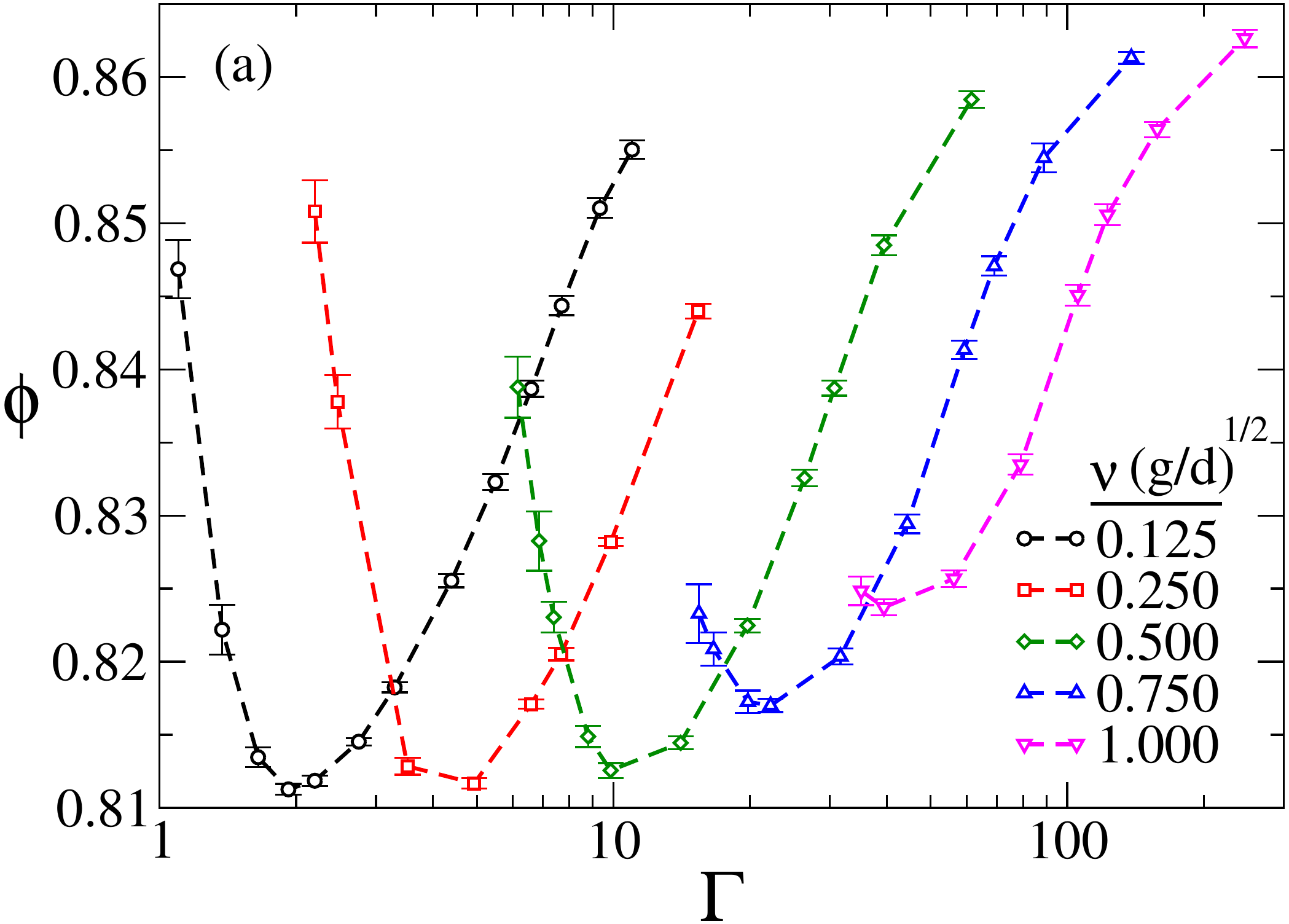}
 \includegraphics[width=0.4\textwidth,angle=-0]{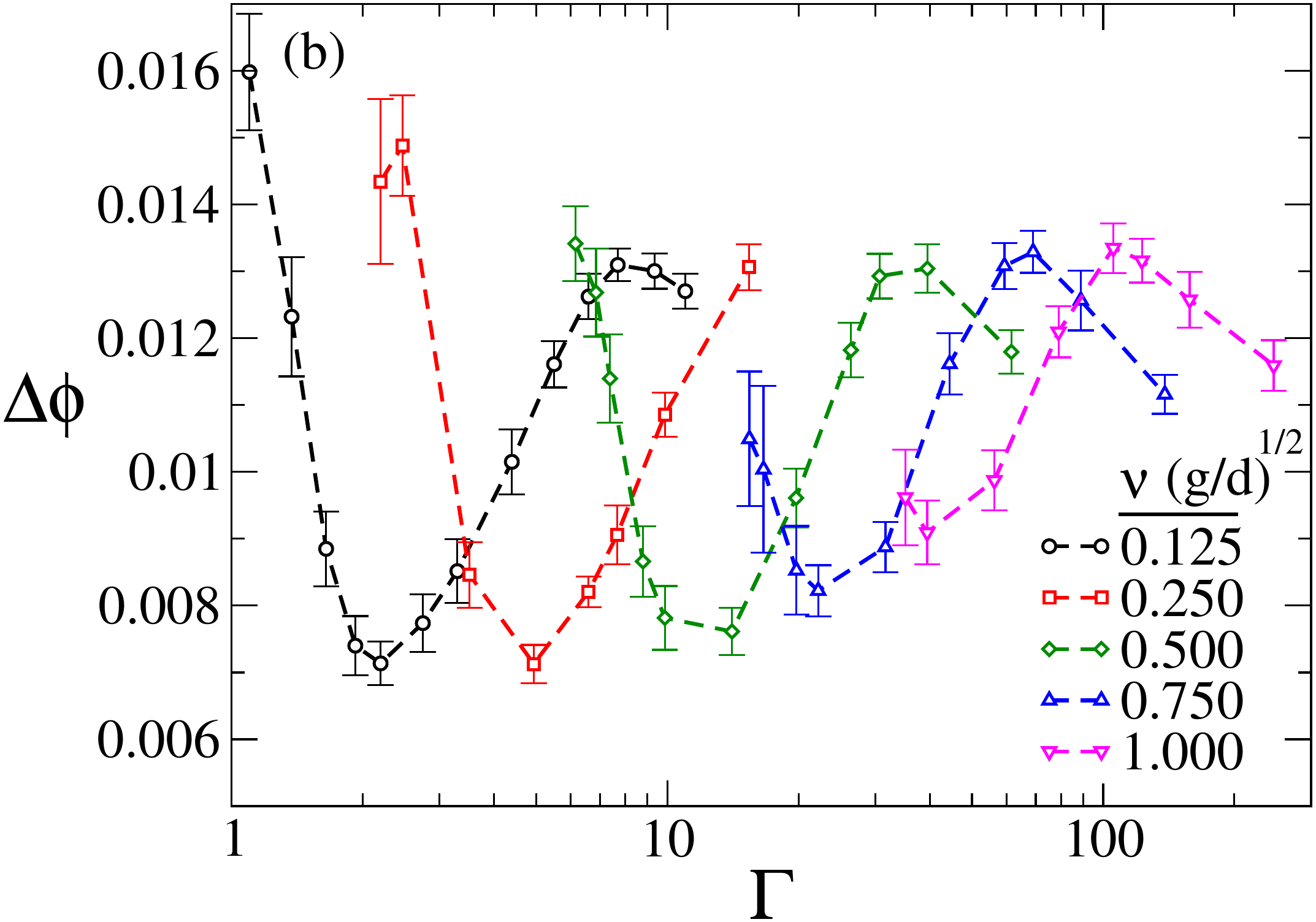}
\end{center}
 \caption{(a) Simulation results for the volume fraction, $\phi$, as a function of the reduced peak acceleration, $\Gamma$, for different frequencies, $\nu$, of the tap pulse. (b) The corresponding volume fraction fluctuation, $\Delta \phi$, as a function of $\Gamma$.}
 \label{fig:phi-vs-Gamma_sim}
\end{figure}

In Fig.~\ref{fig:phi-vs-Gamma_sim}(a), we plot $\phi$ in the steady state as a function of $\Gamma$ for our simulations. Although the fluctuations of $\phi$ are large [see Fig.~\ref{fig:history_simul}], its mean value is well defined with a small confidence interval (see error bars). For low excitations, $\phi$ decreases as $\Gamma$ is increased. However, beyond a certain value $\Gamma_{\rm{min}}$, the packing fraction grows. The same trend is observed if the tap frequency $\nu$ is changed. However, the minimum is deeper for lower $\nu$ and its position $\Gamma_{\rm{min}}$ shifts to larger values of $\Gamma$ as $\nu$ is increased in coincidence with our experiments (see Fig.~\ref{fig:phi-vs-Gamma_expt}). Due to the change in the depth of the $\phi$ minimum in the $\phi$--$\Gamma$ curves, a simple rescaling of $\Gamma$ is unable to collapse the data for different frequencies. However, rescaling $\phi$ and $\Gamma$ with $\phi_{min}$ and $\Gamma_{min}$, respectively, yields very good collapse, even between simulated and experimental data \cite{pugnaloni2010}.

In Fig.~\ref{fig:phi-vs-Gamma_sim}(b), the volume fraction fluctuations, $\Delta \phi$, are plotted as a function of $\Gamma$. As we can see, these fluctuations are non-monotonic, as suggested by our experiments (Fig.~\ref{fig:D_phi-vs-Gamma_expt}). Non-monotonic volume fluctuations have also been reported in Ref.~\cite{schroter}. For $\Delta \phi$, we obtain a minimum and a maximum. We have also observed a maximum in $\Delta \phi$ for values of $\Gamma$ below the ones reported here (see for example Refs. \cite{pugnaloni2010} and \cite{carlevaro}). However, we do not report such low values of $\Gamma$ in this work and focus on tapping intensities that warrant the steady state with a modest number of pulses.

\begin{figure}[htp]
\begin{center}
 \includegraphics[width=0.4\textwidth,angle=0]{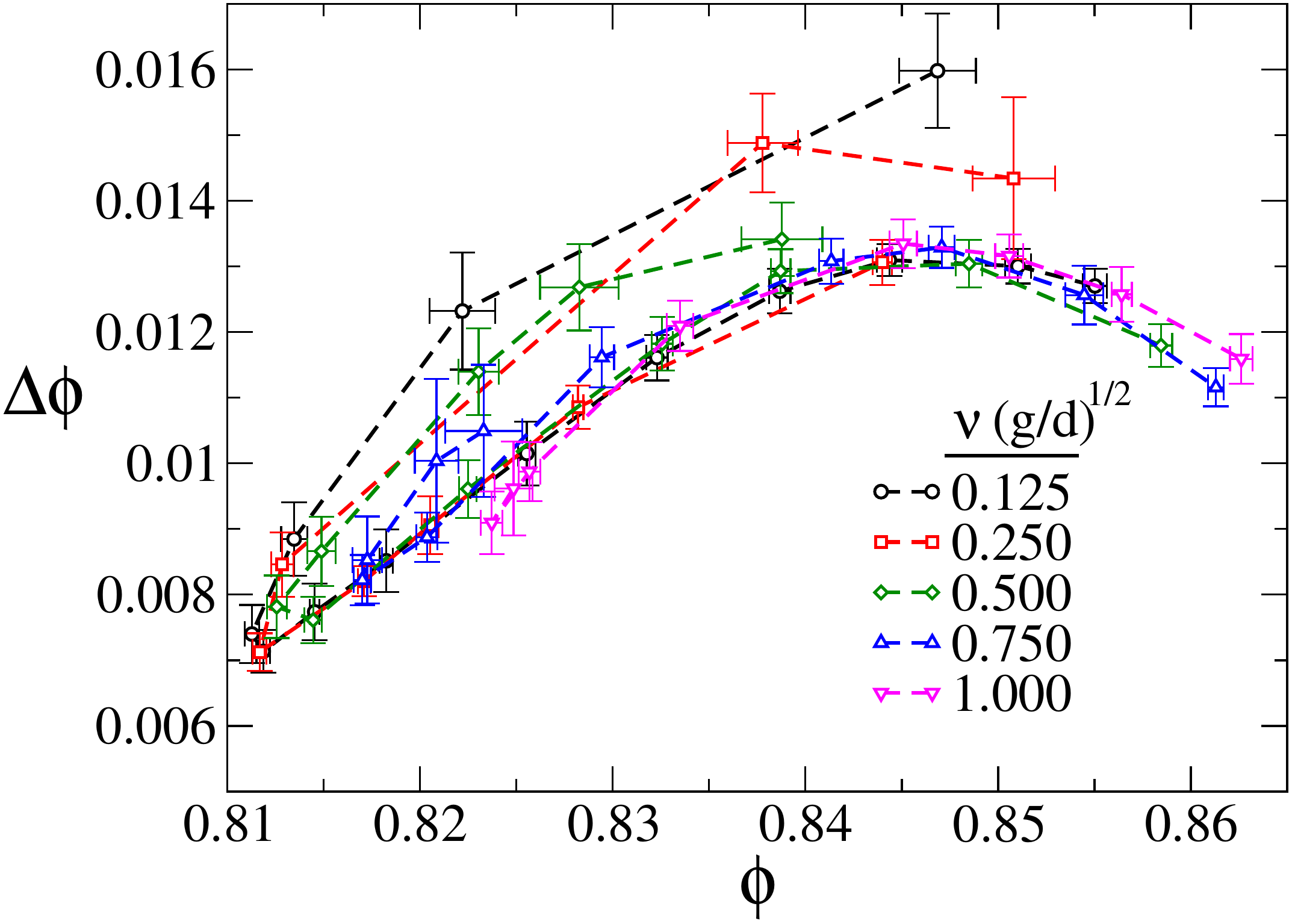}
\end{center}
 \caption{Density fluctuations as a function of $\phi$ for different frequencies of the tap pulse in the simulations.}
 \label{fig:D_phi-vs-phi}
\end{figure}

The value of $\Gamma$, at which the fluctuations display a minimum, coincides with $\Gamma_{min}$, the value at which the minimum packing fraction, $\phi_{min}$, is obtained. The maximum coincides with the inflection point in the $\phi$--$\Gamma$ curve at higher $\Gamma$. Since one expects to find few mechanically stable configurations compatible with a large volume (low $\phi$), it seems reasonable that fluctuations reach a minimum if $\phi$ does so. Similarly, there should be few low-volume, mechanical stable configurations which implies that at high $\phi$ fluctuations should diminish. Hence, a maximum in $\Delta \phi$ should be present at intermediate packing fractions. This is more clearly seen in Fig.~\ref{fig:D_phi-vs-phi} where we plot the fluctuations as a function of the average value of $\phi$.

Figure \ref{fig:D_phi-vs-phi} presents two distinct branches: the lower branch corresponds to $\Gamma > \Gamma_{min}$ and the upper branch to $\Gamma < \Gamma_{min}$. For $\Gamma > \Gamma_{min}$, the fluctuations corresponding to different tap durations collapse, suggesting that such equal-$\phi$, equal-$\Delta \phi$ states might correspond to unique states (below, we will find that this is not the case). For $\Gamma < \Gamma_{min}$, we obtain states of same $\phi$ as some states in the lower branch but presenting larger fluctuations. This is clear evidence that the equal-$\phi$ states of the upper and lower branch are indeed distinct and that other macroscopic variables must be used to distinguish one from the other.

We have assessed a number of other structural descriptors (coordination number, bond order parameter, radial distribution function). In all cases, equal--$\phi$ states from the upper and lower branch of the $\phi$--$\Delta \phi$ curve (Fig.~\ref{fig:D_phi-vs-phi}) present similar values of the structural descriptors with only subtle discrepancies. Although this indicates that the states are not equivalent, it also suggests that such descriptors are not good candidates to form a set of macroscopic variables, along with $V$, to uniquely identify a given steady state. 

In Ref.~\cite{pugnaloni2010}, we have assessed the force moment tensor $\Sigma$ as a good candidate to complete $V$ and $N$ in describing a stationary state. This has been suggested by some theoretical speculations \cite{blumenfeld}. However, some authors prefer to directly replace $V$ by $\Sigma$. Below, we will show that both, $V$ and $\Sigma$, are required for the equilibrium states generated in our simulations. Moreover, we will show that only one of the invariants of $\Sigma$ is necessary (at least in our 2D systems) and that fluctuations of these variables suggest that no other extra macroscopic parameter may be required.

\section{Stress tensor}

Before we focus on the force moment tensor, we will consider the stress tensor, $\sigma$, in order to understand the phenomenology of the force distribution in our tapped granular beds. We recall that $\Sigma$ and $\sigma$ are simply related through $\Sigma \equiv V \sigma$. However, we have to bear in mind that $V$ is not a simple constant since the volume of the system depends, in a nontrivial way, on the shape and intensity of the excitation. 

\begin{figure}[htp]
\begin{center}
 \includegraphics[width=0.4\textwidth,angle=0]{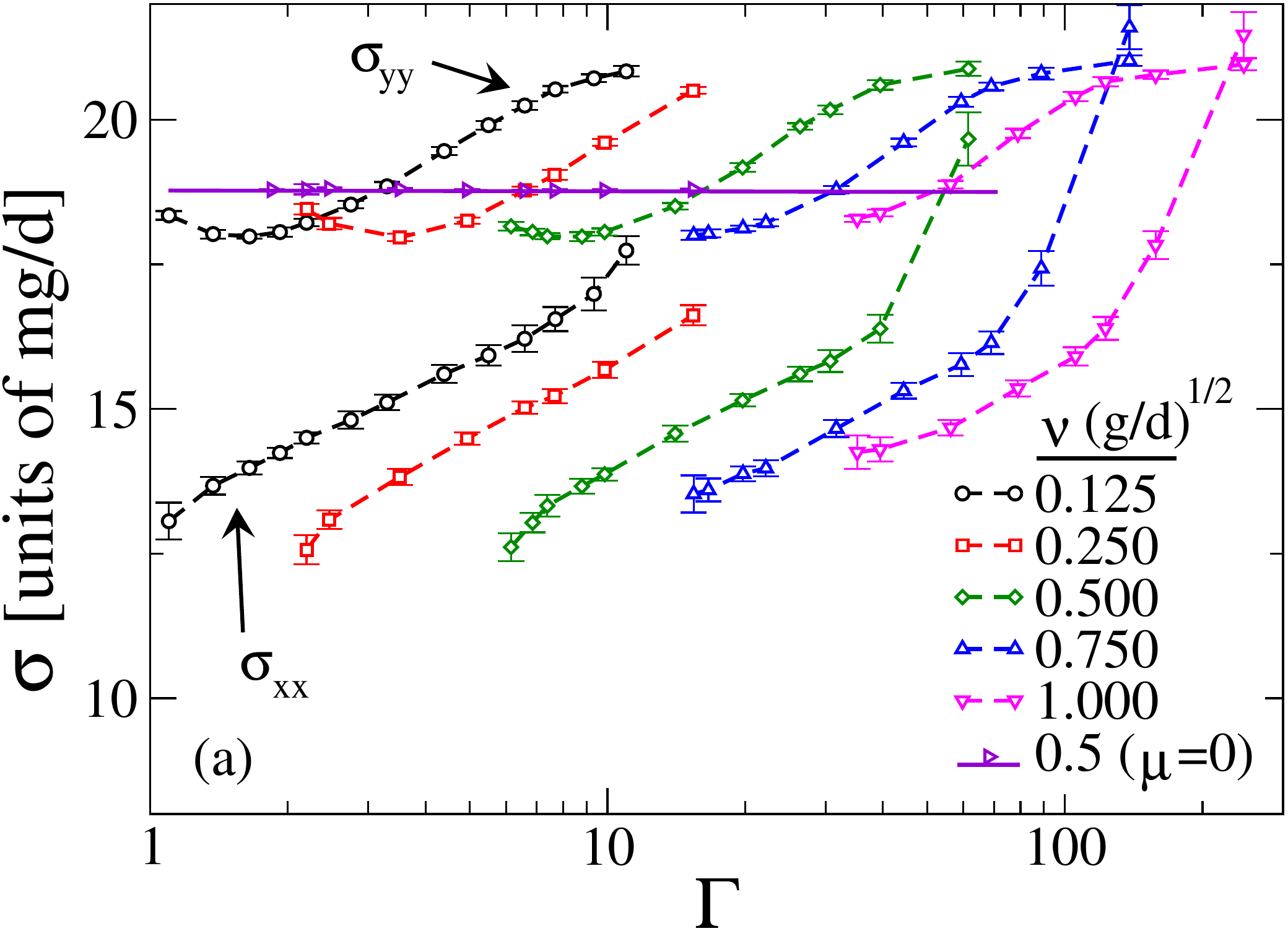}
 \includegraphics[width=0.4\textwidth,angle=0]{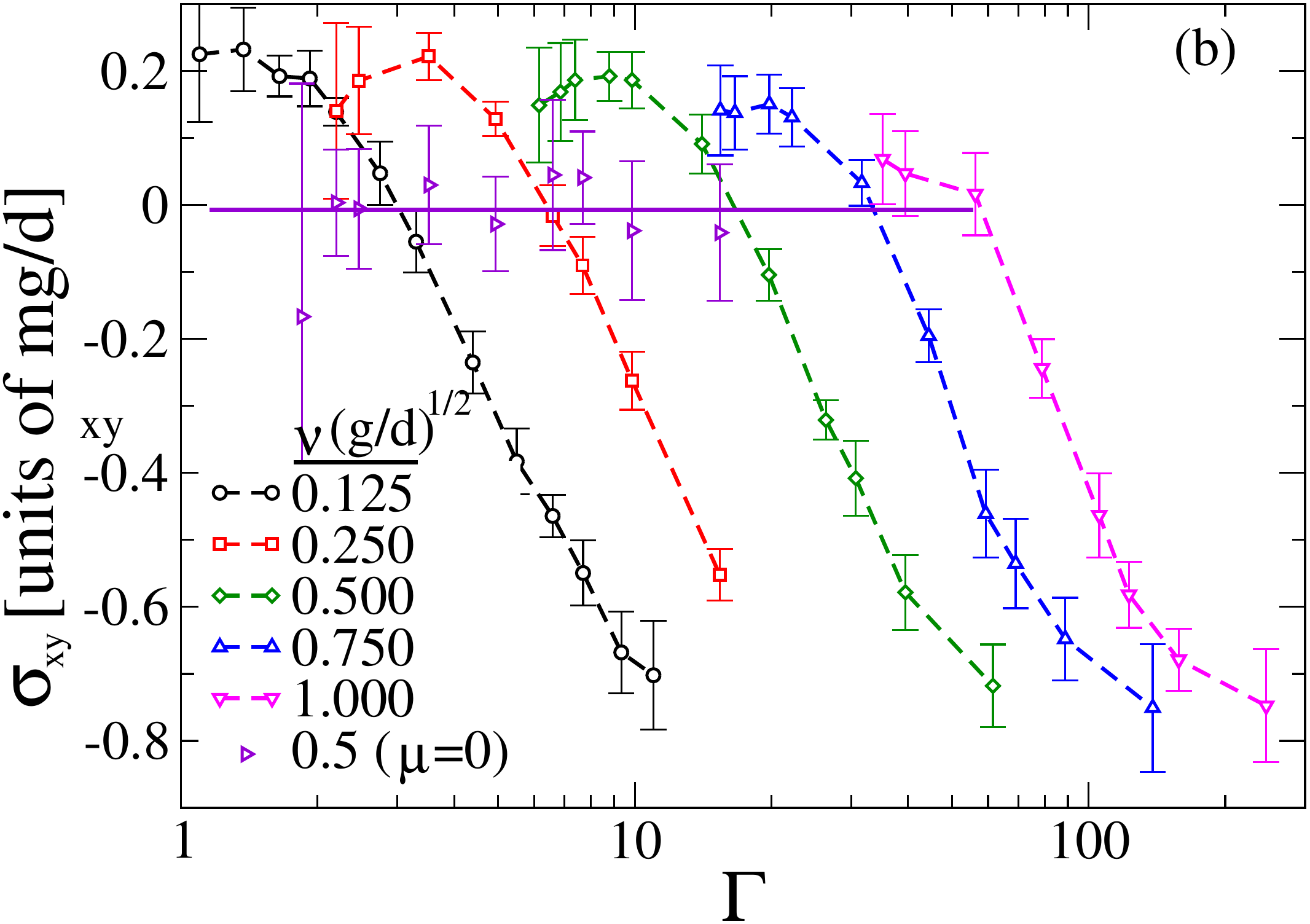}
\end{center}
 \caption{(a) Diagonal components of the stress tensor, $\sigma$, as a function of $\Gamma$ for different frequencies $\nu$ of the tap pulse (the upper set of curves corresponds to $\sigma_{yy}$ and the lower set to $\sigma_{xx}$). (b) Off diagonal component of the stress tensor, $\sigma_{xy}$. The horizontal line corresponds to $\sigma_{yy}$ [in panel (a)] and to $\sigma_{xy}$ [in panel (b)] from simulations of a packing of frictionless disks.}
 \label{fig:sigma_xx_yy-vs-Gamma}
\end{figure}

In Fig.~\ref{fig:sigma_xx_yy-vs-Gamma}, we show the components of $\sigma$ as a function of $\Gamma$ for different $\nu$. As a reference, we show results of our simulations for a frictionless system. In a frictionless system, the shear vanishes and $\sigma_{yy}$ is only determined by the weight of the sample since the Janssen's effect is not present. As we can see, the frictionless sample presents a constant value of $\sigma_{yy}$ for all $\Gamma$. For low $\Gamma$, the frictional samples display values of $\sigma_{yy}$ below the frictionless reference. This is a consequence of the Janssen's effect since part of the weight of the sample is supported by the wall friction. Consequently, in this region, $\sigma_{xy}$ is also positive [see Fig.~\ref{fig:sigma_xx_yy-vs-Gamma}(b)]. However, for each $\nu$, there is a critical value of $\Gamma$, $\Gamma_{shear=0}$. Beyond it, the sample presents an apparent weight above the weight of the packing. In correspondence with this, $\sigma_{xy}$ changes sign and becomes negative. This indicates that, for $\Gamma > \Gamma_{shear=0}$, the frictional walls are not supporting any weight. Rather, they prevent the packing from expansion by a downward frictional force. As $\Gamma$ is increased beyond $\Gamma_{shear=0}$, the packing tends to store most of its stress in the horizontal direction ($\sigma_{xx}$) while $\sigma_{yy}$ eventually saturates. For very intense pulses, the sample expands and lifts off significantly during the tap. When the bed falls back, it creates a very compressed structure with most of the stress transmitted in the lateral directions and the wall friction sustaining the system downwards. It is worth mentioning that $\Gamma_{shear=0}$ is always higher than $\Gamma_{min}$ (see Fig. \ref{fig:phi-vs-Gamma_sim}). 

\section{The force moment tensor master curve}

\begin{figure}[htp]
\begin{center}
 \includegraphics[width=0.4\textwidth,angle=0]{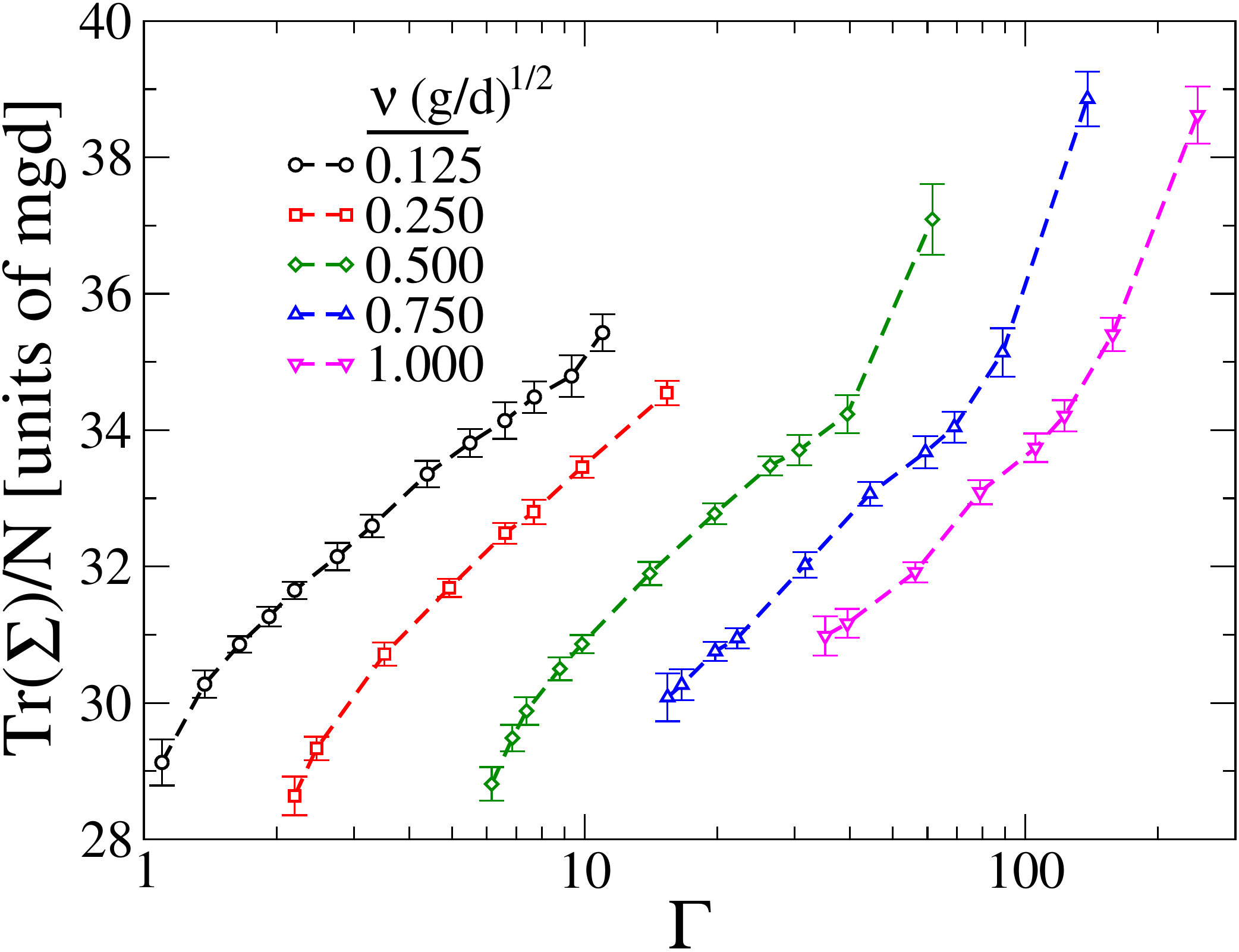}
 \end{center}
 \caption{Trace, $\rm{Tr}(\Sigma)$, of the force moment tensor as a function of $\Gamma$ for different frequencies $\nu$ of the tap pulse.}
 \label{fig:Sigma_tr-vs-Gamma}
\end{figure}

Since the stress is not an extensive parameter, the force moment tensor is generally used to characterize the macroscopic state \cite{blumenfeld,tighe}. Therefore, we will use $\Sigma$ in the rest of the paper. Let us simply remark that since $V$ presents a non-monotonic response to $\Gamma$, the curves in Fig.~\ref{fig:sigma_xx_yy-vs-Gamma} present a somewhat different shape if $\Sigma$ is plotted instead of $\sigma$. Particularly, $\Sigma_{yy}$ does not display a minimum at low $\Gamma$, as the one observed for $\sigma_{yy}$ in frictional disks but a monotonic increase. In Fig.~\ref{fig:Sigma_tr-vs-Gamma}, we show the trace of $\Sigma$ as a function of $\Gamma$. There is a clear monotonic increase of $\rm{Tr}(\Sigma)$ as $\Gamma$ is increased. Moreover, for a given $\Gamma$, if the frequency of the excitation pulse is increased, a significant reduction in the force moment tensor is observed.

\begin{figure}[htp]
\begin{center}
 \includegraphics[width=0.4\textwidth,angle=0]{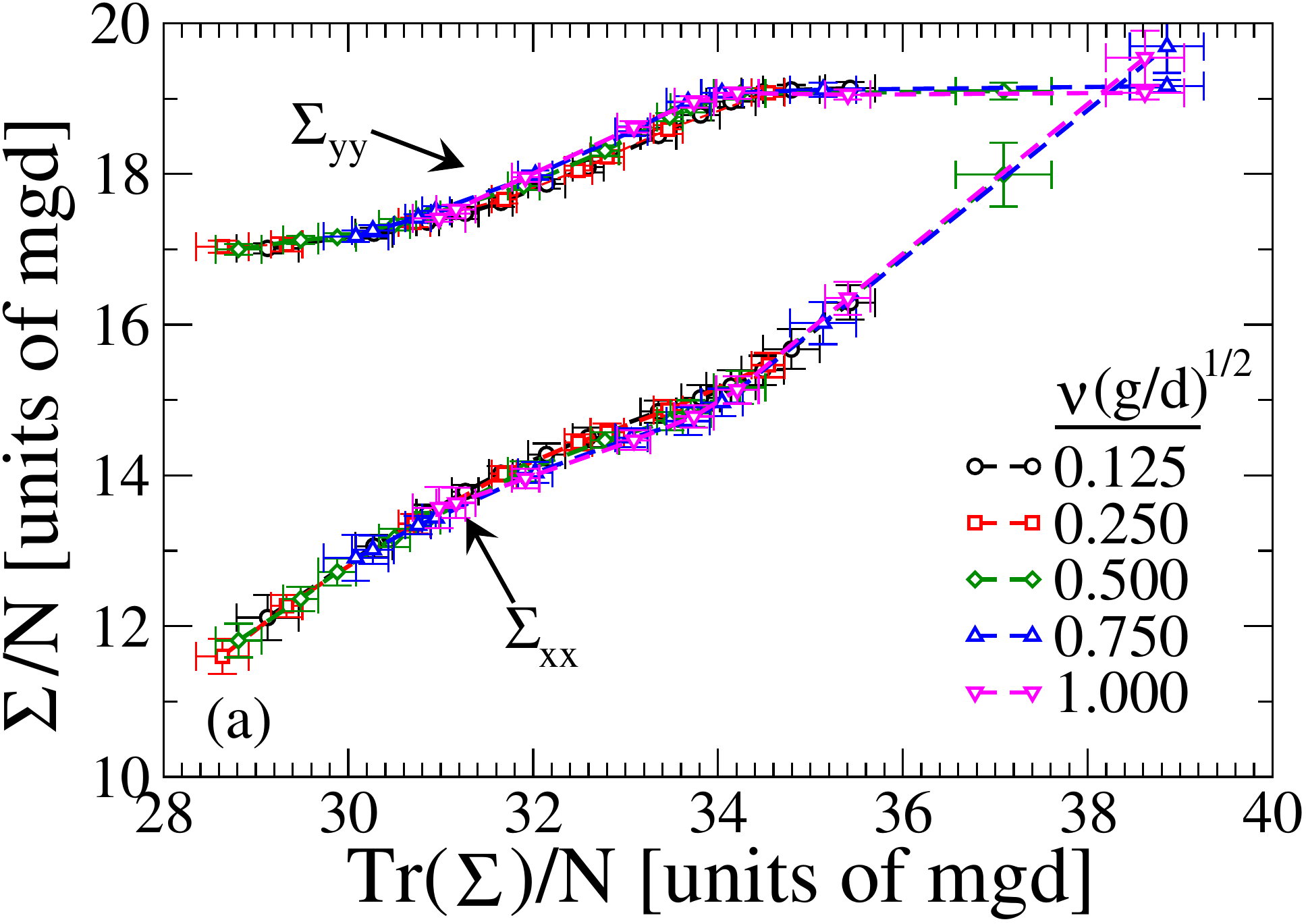}
 \includegraphics[width=0.4\textwidth,angle=0]{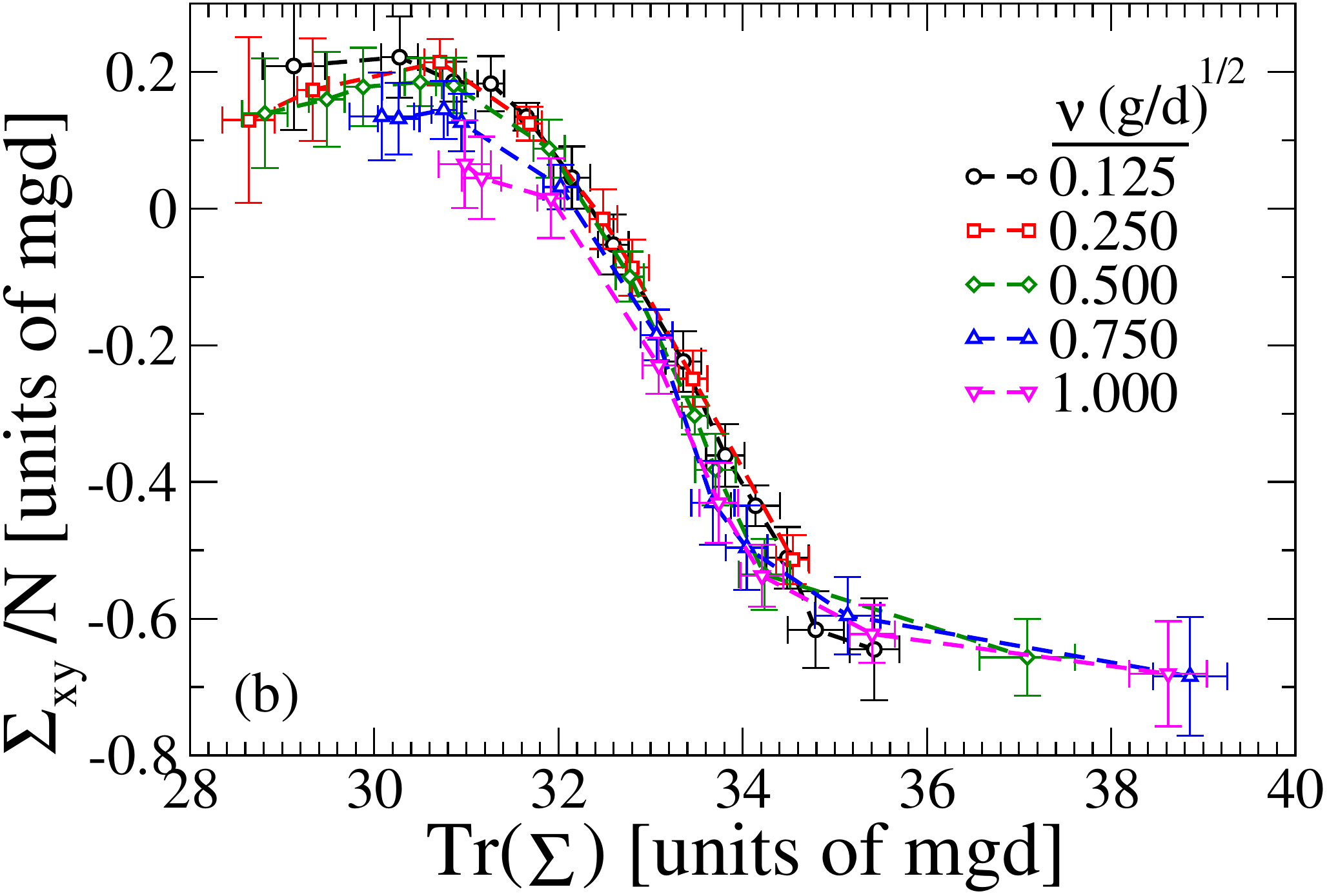}
\end{center}
 \caption{(a) Diagonal components of the force moment tensor, $\Sigma$, as a function of $\rm{Tr}(\Sigma)$ for different frequencies of the tap pulse (the upper set of curves corresponds to $\Sigma_{yy}$ and the lower to $\Sigma_{xx}$). (b) Off diagonal component of the force moment tensor, $\Sigma_{xy}$.}
 \label{fig:Sigma_xx_yy-vs-trace_Sigma}
\end{figure}

In Fig. \ref{fig:Sigma_xx_yy-vs-trace_Sigma}, we plot the components of $\Sigma$ as a function of its trace for all the steady states generated. We can see that all data for different $\Gamma$ and $\nu$ collapse into three master curves. This indicates that if two equilibrium states present the same $\rm{Tr}(\Sigma)$, all the components of $\Sigma$ are also equal. Here, we point out a relevant piece of information that will be discussed in the next section. Two states may present equal force moment tensor but differ in volume. This means that many points collapsing in Fig. \ref{fig:Sigma_xx_yy-vs-trace_Sigma} correspond to states of different $\phi$. Therefore, at equilibrium, irrespective of the structure of the sample, two states with the same trace in $\Sigma$ will present equal $\Sigma$.

In a liquid at equilibrium, the stress tensor is diagonal and all elements along the diagonal are equal. This hydrostatic property allows us to know the full stress tensor if we only know the hydrostatic pressure (i.e., if we only know the trace of the tensor). In our granular samples, the force moment tensor can also be known if the trace is known. However, the shape of the tensor in static packings under gravity is defined by the three master curves of Fig.~\ref{fig:Sigma_xx_yy-vs-trace_Sigma}. 

To our knowledge, there is no previous speculation that this property must hold for static granular packings. A more detailed study on the extent of this commonality of the shape of the force moment tensor will be pursued in a future paper \cite{luis2}. However, we show some suggestive preliminary results below. 

\begin{figure}[htp]
\begin{center}
 \includegraphics[width=0.40\textwidth,angle=0]{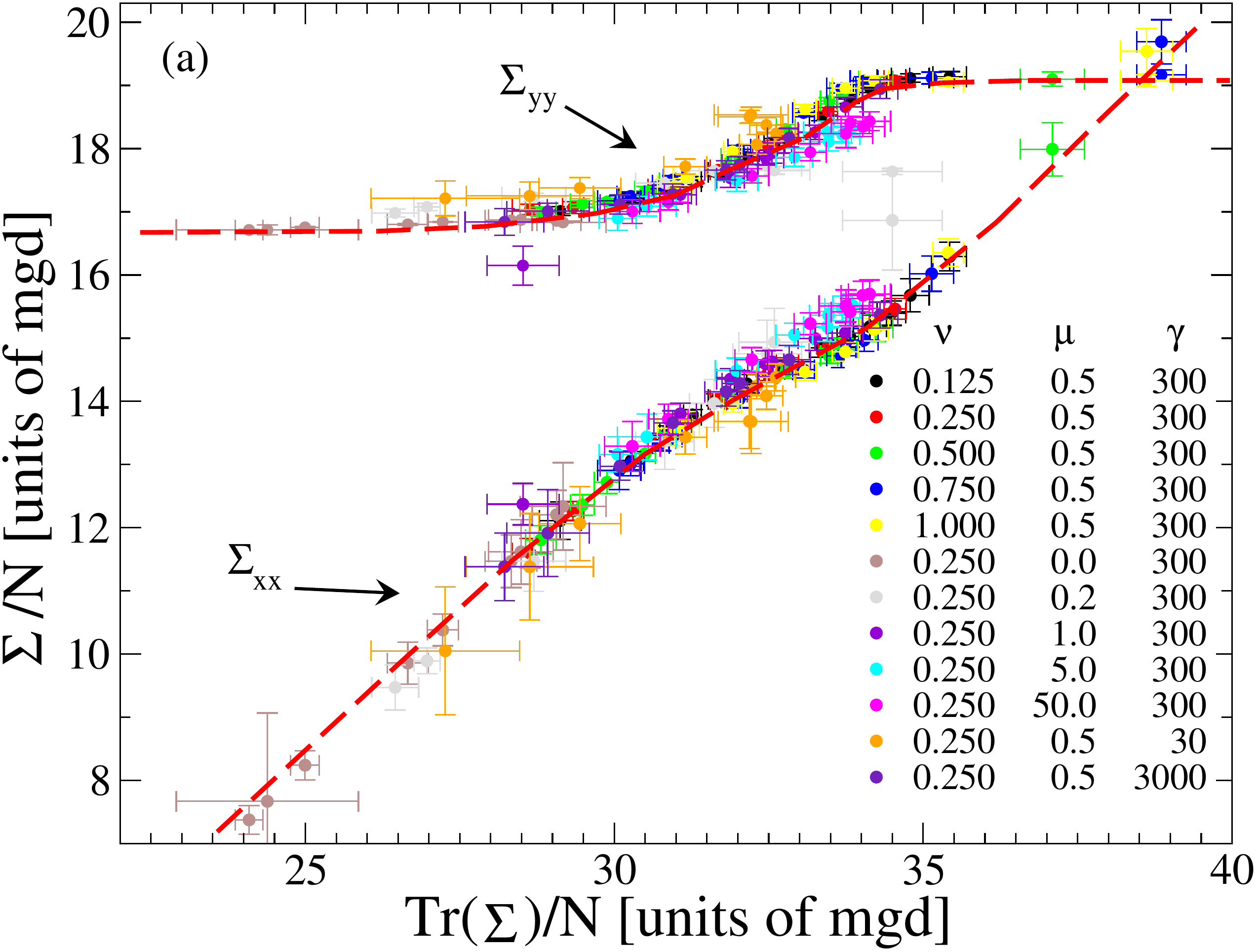}
 \includegraphics[width=0.40\textwidth,angle=0]{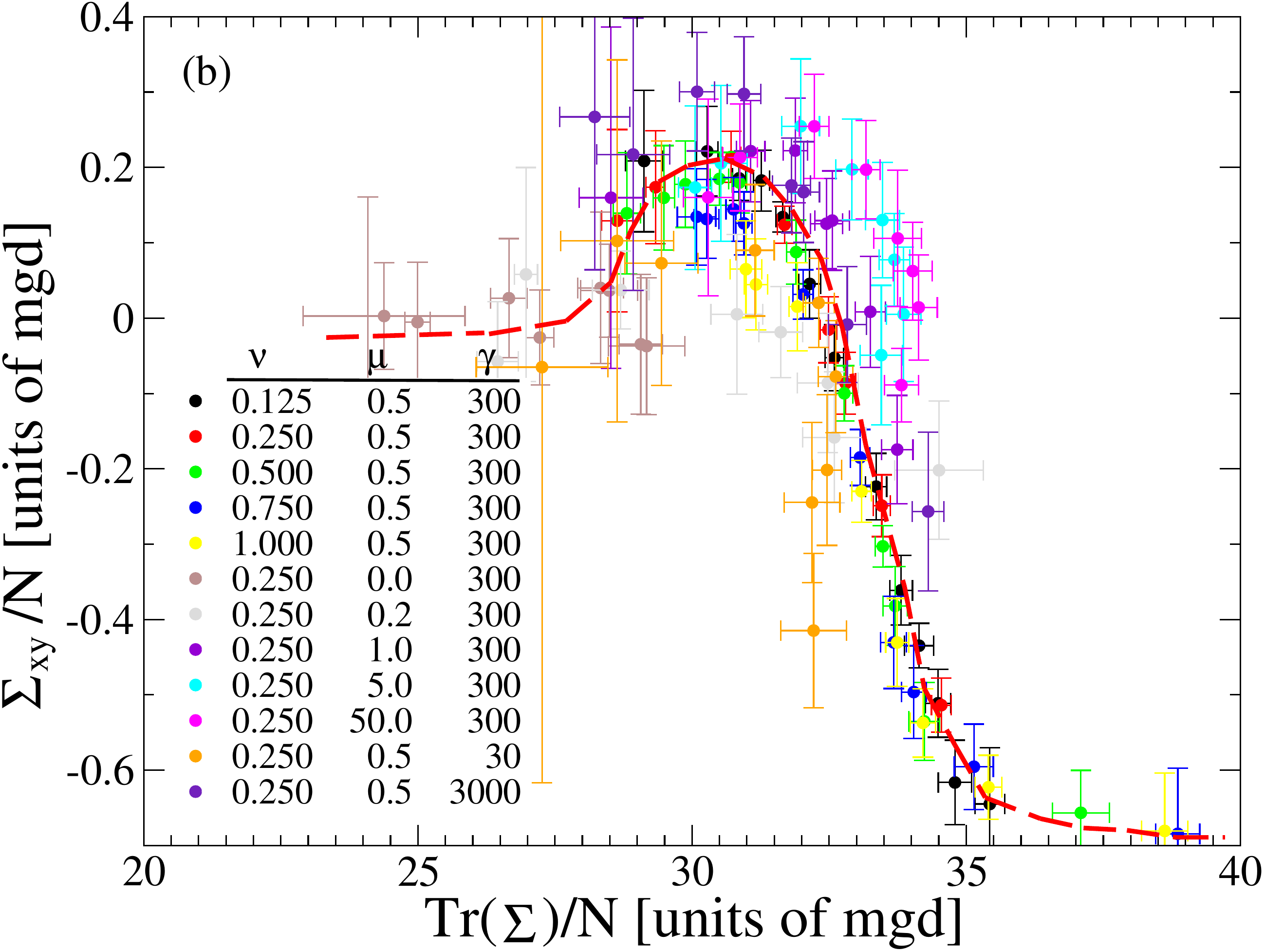}
\end{center}
 \caption{(a) Diagonal components of force moment tensor, $\Sigma$, as a function of $\rm{Tr}(\Sigma)$ for different frequencies of the tap pulse, different friction coefficient and different restitution (the upper set of curves corresponds to $\Sigma_{yy}$ and the lower to $\Sigma_{xx}$). (b) Off diagonal component of the force moment tensor, $\Sigma_{xy}$. The dashed lines are only a guide to the eyes.}
 \label{fig:Sigma_xx_yy-vs-traza_Sigma_friction_restitution}
\end{figure}

In Fig.~\ref{fig:Sigma_xx_yy-vs-traza_Sigma_friction_restitution}, we show the components of $\Sigma$ as a function of $\rm{Tr}(\Sigma)$ for a range of samples of different materials, for different tapping intensities, and for different tapping frequencies. As we can see, there is reasonable collapse of the data onto the same three master curves shown in Fig.~\ref{fig:Sigma_xx_yy-vs-trace_Sigma}. This is an indication that these master curves may be universal and enclose a rather fundamental underlying property (inaccessible to us at this point) of static granular beds.

\section{Force moment tensor fluctuations}

\begin{figure}[htp]
 \begin{center}
\includegraphics[width=0.4\textwidth,angle=0]{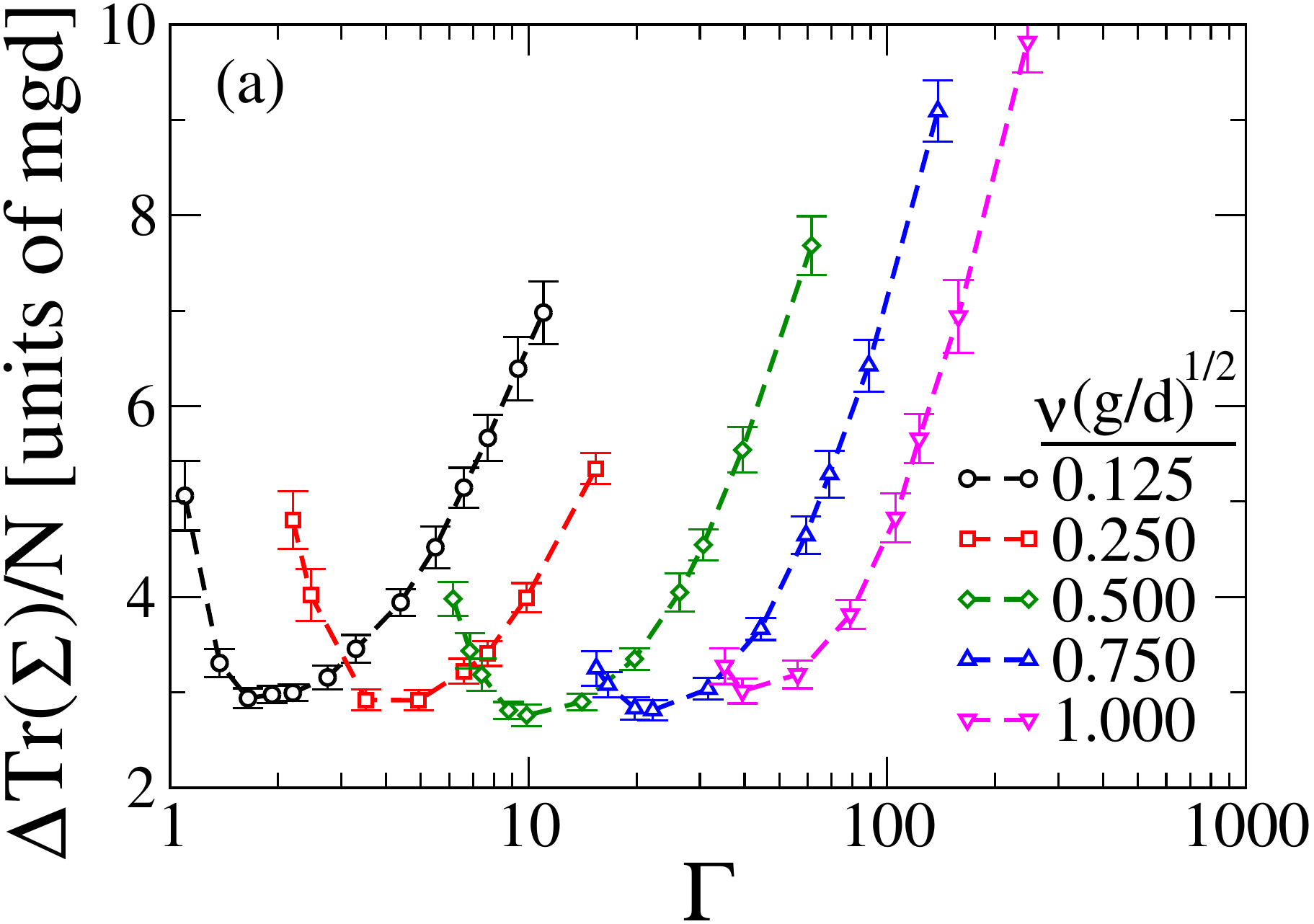}
 \includegraphics[width=0.4\textwidth,angle=0]{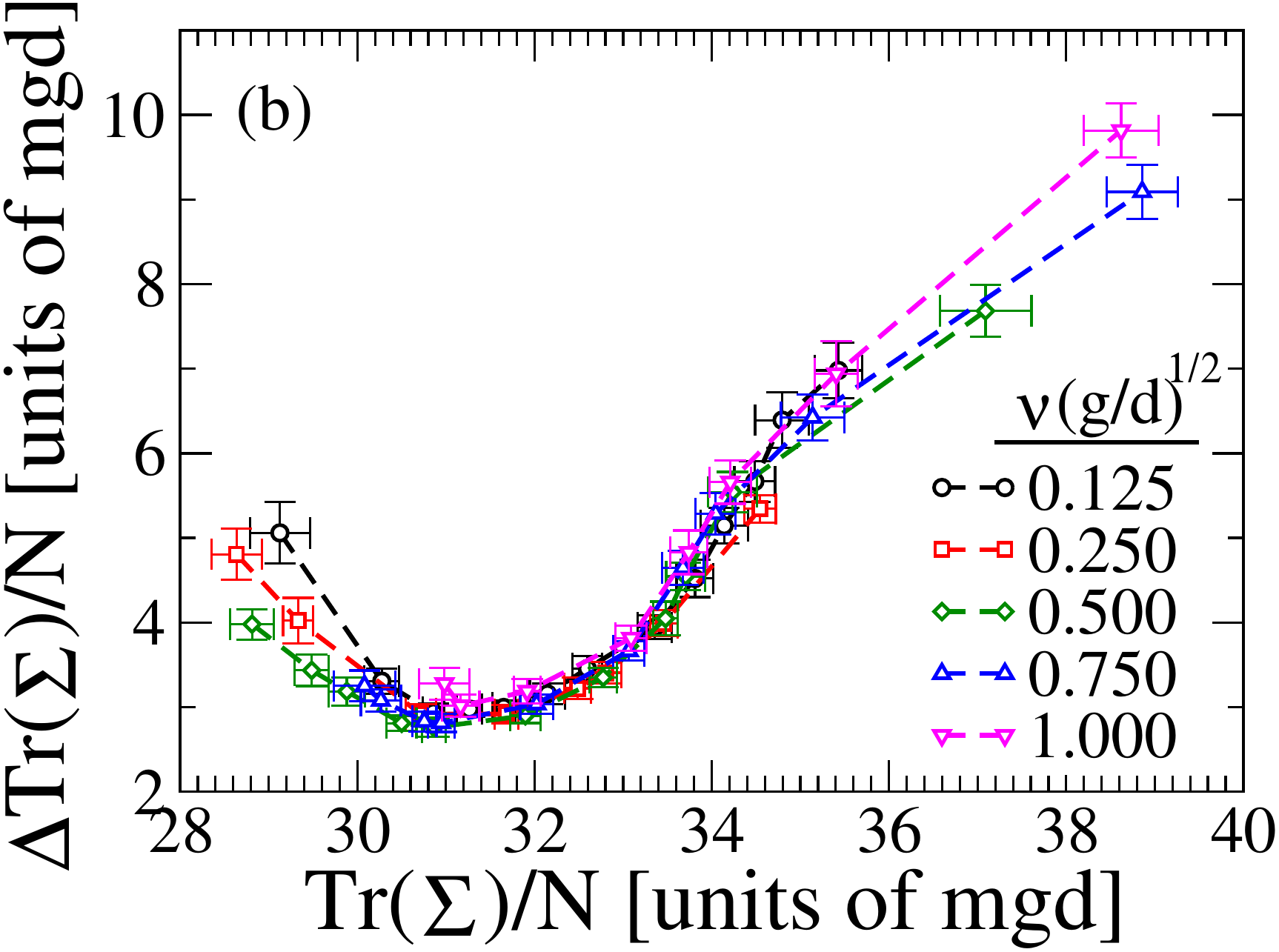}
  \end{center}
\caption{(a) Fluctuations of the trace of the force moment tensor as a function of $\Gamma$ for different frequencies of the tap pulse. (b) Fluctuations of the trace of the force moment tensor as a function of $\rm{Tr}(\Sigma)$.}
 \label{fig:D_trace_Sigma-vs-Gamma}
\end{figure}

Since we have shown in the previous section that only the trace of $\Sigma$ suffices (along with the master curves) to describe the full force moment tensor, we will now focus on this invariant and its fluctuations. In Fig.~\ref{fig:D_trace_Sigma-vs-Gamma}, the fluctuations of $\rm{Tr}(\Sigma)$ are plotted as a function of $\Gamma$. We obtain a single minimum, in contrast with the minimum and maximum observed in $\Delta \phi$. Interestingly, the states with minimum $\Delta \rm{Tr}(\Sigma)$ correspond to the states where the minimums $\phi$ and $\Delta \phi$ are reached for each $\nu$. However, unlike $\Delta \phi$, the depth of the minimum in $\Delta \rm{Tr}(\Sigma)$ is fairly independent of $\nu$. It is unclear why the force moment fluctuations should present a minimum. Provided that the minimum of $\Delta \rm{Tr}(\Sigma)$ coincides with the minimum of $\Delta \phi$, it can be speculated that a reduced number of geometric configurations can accommodate a limited number of force configurations. We have seen that all individual components of $\Sigma$ present the same minimum in their fluctuations, however, the actual values in $\Delta\rm{Tr}(\Sigma$) are dominated by $\Delta\Sigma_{xx}$ which takes values five times larger than $\Delta\Sigma_{yy}$.

If we plot $\Delta\rm{Tr}(\Sigma)$ in terms of the average value $\rm{Tr}(\Sigma)$ [see Fig.~\ref{fig:D_trace_Sigma-vs-Gamma}(b)], we can see that the curves collapse on top of each other over a wide range of $\rm{Tr}(\Sigma)$. However, some deviations are apparent at very low and very high forces. Although the fair collapse of the curves suggests that states of equal-$\Sigma$ may correspond to the same equilibrium states, we will see in the next section that many of these equilibrium states are distinguishable through the volume. The inflection observed at high $\rm{Tr}(\Sigma)$ corresponds to the change of regime observed in Fig. \ref{fig:sigma_xx_yy-vs-Gamma}(a) where the vertical stress saturates and most of the contact forces are directed in the $x$--direction.

\section{The thermodynamic phase space}

\begin{figure}[htp]
\begin{center}
 \includegraphics[width=0.4\textwidth,angle=0]{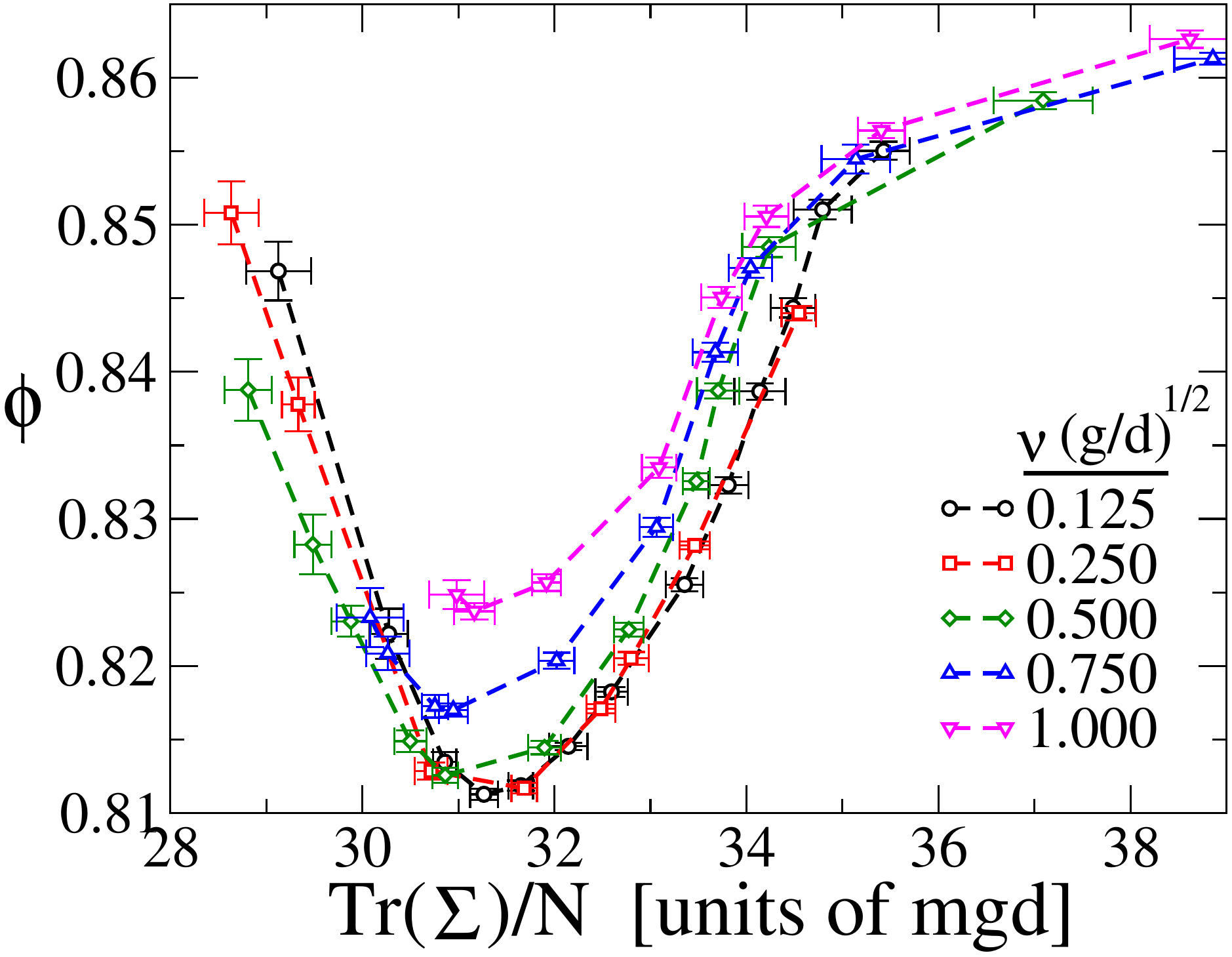}
\end{center}
 \caption{Phase space $\phi$--$\rm{Tr}(\Sigma)$. Loci visited in the simulations for different frequencies of the tap pulse.}
 \label{fig:phi-vs-trace_Sigma}
\end{figure}

As we have suggested, the mean volume of static granular samples is not sufficient to describe the equilibrium state since states of equal-$\phi$ may present distinct fluctuations. On the other hand, the force moment tensor seems to be able to serve as a standalone descriptor since states of equal $\Sigma$ do generally present the same $\Sigma$ fluctuations. However, states of equal-$\Sigma$ may present different volumes. In Fig.~\ref{fig:phi-vs-trace_Sigma}, we plot the loci of the equilibrium states generated in our simulations in a hypothetical $\phi$--$\Sigma$ thermodynamic phase space. As we can see, states of equal $V$ but different $\Sigma$ are obtained as well as states of equal $\Sigma$ and different $V$.
 
\begin{figure}[htp]
\begin{center}
 \includegraphics[width=0.4\textwidth,angle=0]{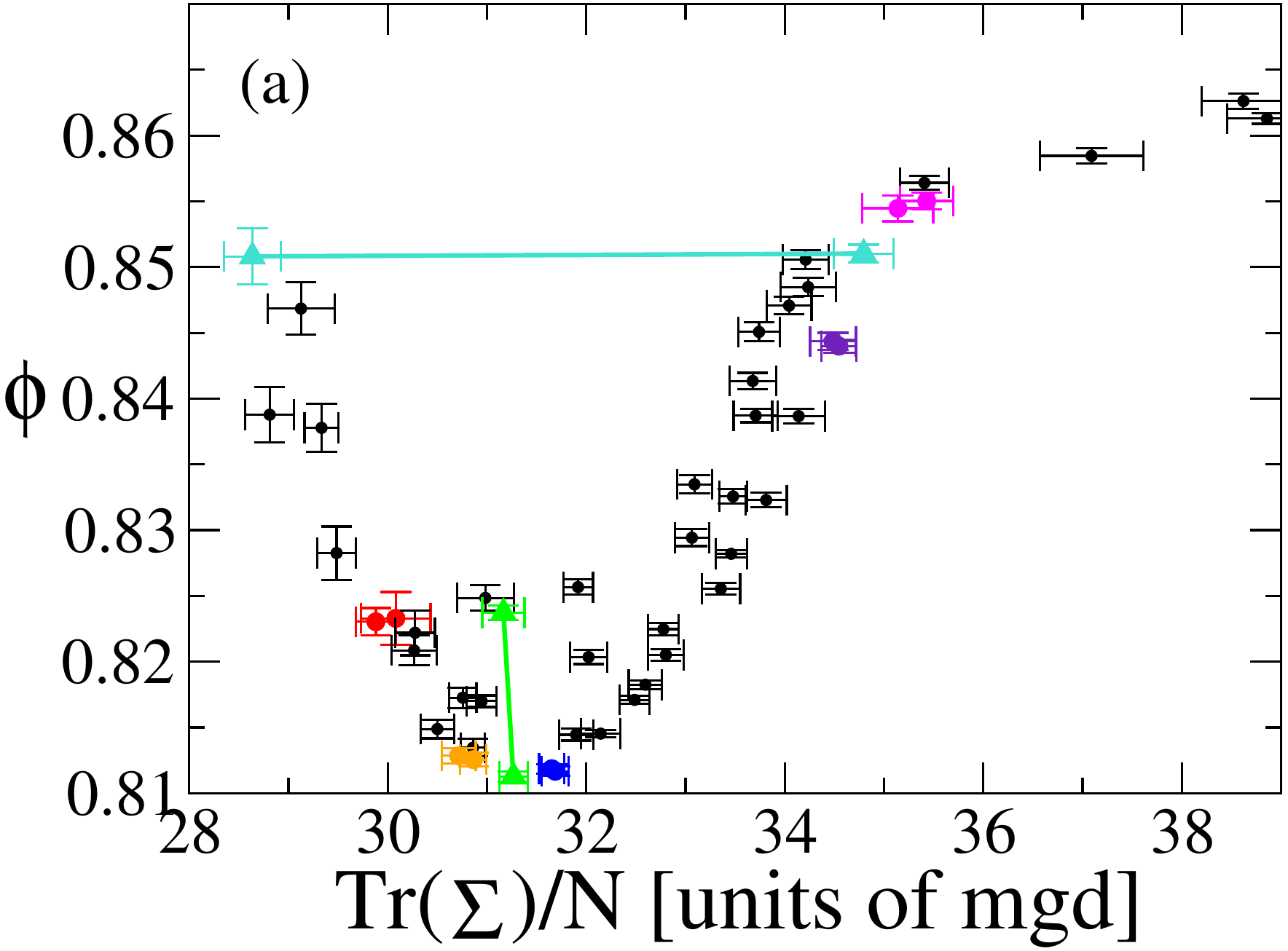}
 \includegraphics[width=0.4\textwidth,angle=0]{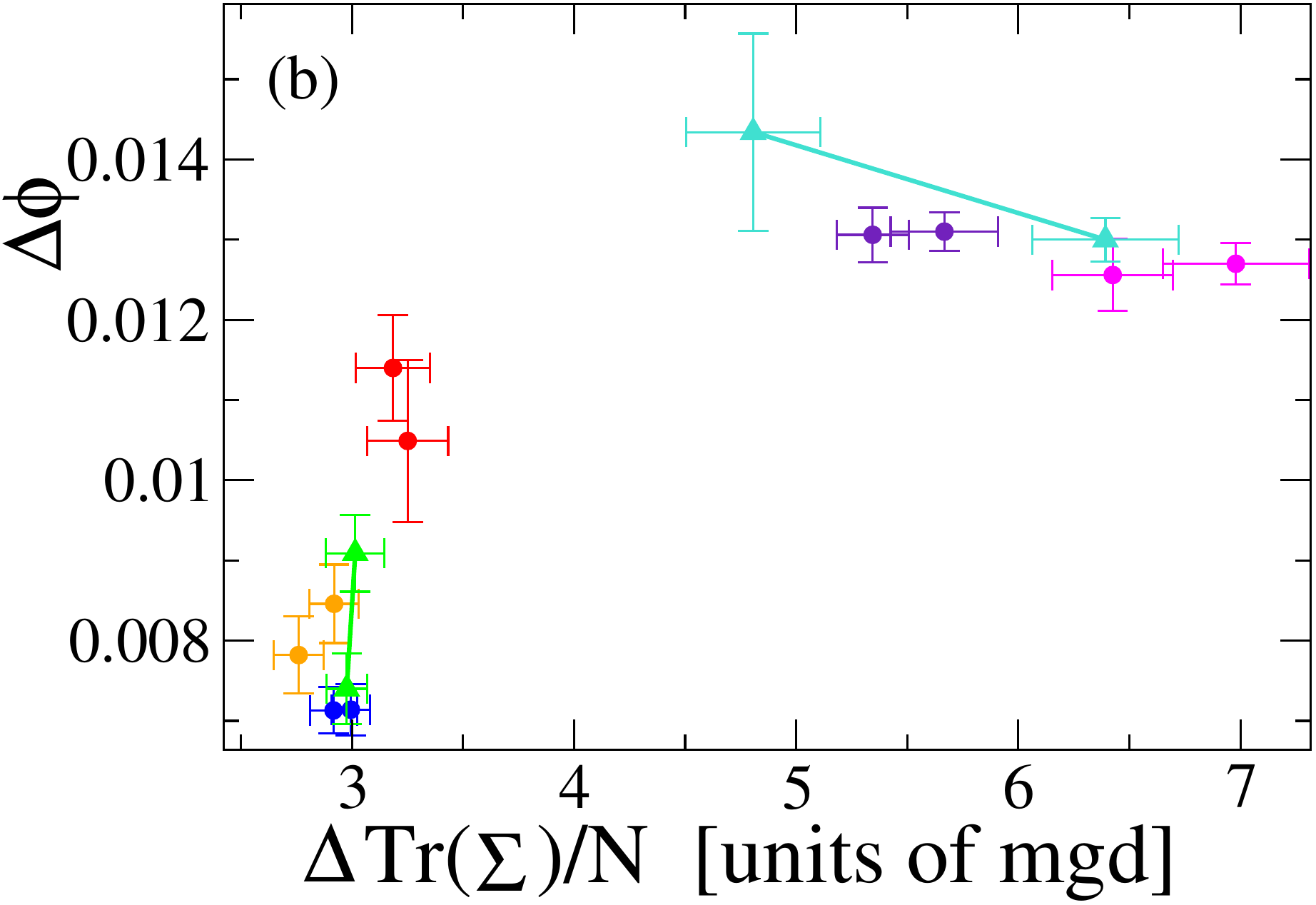}
\end{center}
 \caption{(a) Phase space $\phi$--$\rm{Tr}(\Sigma)$. (b) Fluctuations of the state variables. Selected neighboring states are colored in pairs for comparison. Distant states of equal $V$ or $\rm{Tr}(\Sigma)$ are joined by thick lines.}
 \label{fig:match2}
\end{figure}

We can ask now if these two state variables suffice to fully describe the equilibrium states. A hint that this may be the case is given by the fact that states generated with different $\Gamma$ and $\nu$ but that correspond to the same state in the $\phi$--$\Sigma$ plot display the same fluctuations of these variables. In Fig.~\ref{fig:match2}, we have highlighted some pairs of neighboring states. We can see that such states do also present similar fluctuations [Fig.~\ref{fig:match2}(b)]. In contrast, states which are distant in the $\phi$--$\Sigma$ plot present distinct fluctuations even if they correspond to an equivalent mean volume or an equivalent mean force moment tensor (see states joined by solid lines in Fig.~\ref{fig:match2}).

Let us point out here, that the sole coincidence of fluctuations in the macroscopic variables is not a rigorous proof that the set of chosen variables is a full complete set of thermodynamic parameters. Future explorations of these systems may confirm or disprove that $N$, $V$ and $\rm{Tr}(\Sigma)$ are a full set of macroscopic, extensible variables able to describe all equilibrium states. Meanwhile, it is clear that for moderate tapping intensities, around which the minimum in $\phi$ is observed, the approximation of a simple $NV$ or $N\Sigma$ ensemble is not warranted in view of the large discrepancies between the curves generated with different pulse frequencies in Fig.~\ref{fig:phi-vs-trace_Sigma}.

\section{Concluding remarks}

We have studied steady states of mechanically stable granular samples driven by tap like excitations. We have varied the external excitation by changing both, the pulse amplitude and the pulse duration. We have considered the macroscopic extensive variables $V$ (volume) and $\Sigma$ (force moment tensor), and their fluctuations. From the results, we can draw the following conclusions:

\begin{itemize}
 \item There seems to be a rather robust set of master curves for $\Sigma_{\alpha\beta}$ which implies that the knowledge of $\rm{Tr}(\Sigma)$ suffices to infer the other components of the force moment tensor.
 \item The equilibrium states cannot be only described by $V$ or $\Sigma$, apart from the number of particles $N$.
 \item The equilibrium states seem to be well described by the set $NV\rm{Tr}(\Sigma)$.
\end{itemize}

There exists a number of points to be considered in view of these findings. Here, we mention a few that may serve as starting points for future directions of research:

\begin{itemize}
 \item What is the extent to which the $\Sigma$ master curves are applicable? Is this dependent on the dimensionality of the system, the excitation procedure, the chosen contact force law, etc? 
 \item What is the dynamics during a single pulse that leads to the appearance of the $\phi$ minimum? Is this minimum present in states generated with other types of pulses like fluidization or shear? The ubiquity of this minimum in simulation models \cite{pugnaloni2008,gago} suggests that it might be found in numerous conditions.
 \item Are the fluctuations shown in Figs.~\ref{fig:D_phi-vs-phi} and \ref{fig:D_trace_Sigma-vs-Gamma} the definitive phenomenological equation of states? Other authors have so far found monotonic density fluctuations \cite{ciamarra} or concave up density fluctuations \cite{schroter}. 
 \item How much of the $\phi$--$\rm{Tr}(\Sigma)$ plane can be explored by changing material properties?
 \item Are there other excitation protocols (such as shearing) that may give rise to steady states that are thermodynamically equivalent to the ones obtained by tapping?
 \item Is it possible to construct a phenomenological entropy function from the equations of states (Figs.~\ref{fig:D_phi-vs-phi} and \ref{fig:D_trace_Sigma-vs-Gamma}) by simple integration of a Gibbs--Duhem-like equation? Let us bear in mind that Fig.~\ref{fig:D_phi-vs-phi} is multiply valued.
\end{itemize}

It is worth stressing that if two ensembles generated by arbitrary excitation protocols ---such as tapping or shearing--- happen to present the same mean values (and fluctuation) for all macroscopic variables, then such macroscopic states should be considered thermodynamically identical. However, it may be the case that a given protocol produces a narrow range of macroscopic states that can be eventually described with a reduced set of macroscopic variables.

\begin{acknowledgements}
LAP acknowledges discussions with Massimo Pica Ciamarra. JD acknowledges a scholarship of the FPI program from Ministerio de Ciencia e Innovación (Spain). This work has been financially supported by CONICET (Argentina), ANPCyT (Argentina), Project No. FIS2008-06034-C02-01 (Spain) and PIUNA (Univ. Navarra). 
\end{acknowledgements}


\begin{thebibliography}{50}
\bibitem{expansion} Notice however, that this expansion parameter is not based on the properties of the mechanical excitation itself but on the sample response to it.
\bibitem{equilibrium} According to Callen \cite{callen}, a system is at equilibrium if Thermodynamics holds for such state.
\end{thebibliography}
\end{document}